\documentclass[prl,twocolumn,showpacs,amsmath,amssymb]{revtex4-1}

\usepackage[dvipdfmx]{graphicx}

\usepackage{bm}
\usepackage{color}

\newcommand{\beq}{\begin{equation}}
\newcommand{\eeq}{\end{equation}}
\newcommand{\bea}{\begin{eqnarray}}
\newcommand{\eea}{\end{eqnarray}}
\newcommand{\bec}{\begin{center}}
\newcommand{\enc}{\end{center}}
\newcommand{\bfr}{\begin{flushright}}
\newcommand{\efr}{\end{flushright}}
\newcommand{\la}{\langle}
\newcommand{\ra}{\rangle}

\newcommand{\om}{\omega}
\newcommand{\kap}{\kappa}
\newcommand{\gam}{\gamma}
\newcommand{\s}{\sigma}

\newcommand{\Om}{\Omega}

\newcommand{\tc}{\widetilde{c}}

\newcommand{\cH}{{\cal H}}
\newcommand{\cL}{{\cal L}}

\begin{document}

\title{Nonclassical photon number distribution in a superconducting cavity
\\under a squeezed drive}

\author{S. Kono$^{1}$}
\author{Y. Masuyama$^{1}$}
\author{T. Ishikawa$^{1}$}
\author{Y. Tabuchi$^{1}$}
\author{R. Yamazaki$^{1}$}
\author{K. Usami$^{1}$}
\author{K. Koshino$^{2}$}
\author{Y. Nakamura$^{1,3}$}%

\affiliation{$^{1}$Research Center for Advanced Science and Technology (RCAST), The University of Tokyo, Meguro-ku, Tokyo 153-8904, Japan}
\affiliation{$^{2}$College of Liberal Arts and Sciences, Tokyo Medical and Dental University, Ichikawa, Chiba 272-0827, Japan}
\affiliation{$^{3}$Center for Emergnent Matter Science (CEMS), RIKEN, Wako, Saitama 351-0198, Japan}

\date{\today}

\begin{abstract}
A superconducting qubit in the strong dispersive regime of circuit quantum electrodynamics is a powerful probe for microwave photons in a cavity mode.
In this regime, a qubit excitation spectrum is split into multiple peaks, with each peak corresponding to an individual photon number in the cavity (discrete ac Stark shift).
Here, we measure the qubit spectrum in a cavity that is driven continuously with a squeezed vacuum generated by a Josephson parametric amplifier. 
By fitting the obtained spectrum with a model which takes into account the finite qubit excitation power,
we determine the photon number distribution, which reveals an even-odd photon number oscillation and quantitatively fulfills Klyshko's criterion for nonclassicality.
\end{abstract}


\maketitle

Advancement of the superconducting quantum circuit technologies~\cite{scq} and the concept of circuit quantum electrodynamics (QED)~\cite{cqed} have led to the emergence of microwave quantum optics, enabling us to generate and characterize nonclassical states of electromagnetic fields in the microwave domain.

A squeezed vacuum is one of the most widely studied nonclassical states as a resource in quantum technologies,
such as computation, communication and metrology~\cite{qsq}. 
In microwave quantum optics, a squeezed vacuum is conveniently generated by degenerated parametric down conversion in a Josephson parametric amplifier (JPA) based on the nonlinearity of Josephson junctions~\cite{ozn,asq}.  
Characterizations of such states propagating in a waveguide have been realized by measuring the quadrature amplitudes with a homodyne technique with the aid of a JPA~\cite{qsm} or a cryogenic HEMT amplifier~\cite{pec,sfj}.  
JPAs and related circuits are also used to generate and characterize two-mode squeezing in spatially or spectrally separated propagating modes~\cite{odc, ots, tmc, gem, jtp}. 
More recently, it has been shown that a squeezed vacuum injected in a cavity induces nontrivial effects to the relaxations of a qubit~\cite{rrd,rfa} and a spin ensemble~\cite{mrs}. 
In the Fock basis, on the other hand, a squeezed vacuum displays another feature of the nonclassicality, i.e., the photon number distribution composed of only even photon numbers~\cite{eoo}. 
In the optical domain, direct observations of photon number distribution using a photon-number-resolving detector were reported~\cite{hnp,udd}.
In the microwave domain, however, because of the smallness of the energy of a single photon, photon counting in a propagating mode is still a challenging task, while a few realizations of microwave single-photon detectors have been reported~\cite{mpc,spd,rcr}.

Here, we report the measurement of the photon number distribution of a squeezed vacuum continuously injected into a cavity containing a superconducting qubit.
In the strong dispersive regime of the circuit-QED architecture, the spectrum of a superconducting qubit is split into multiple peaks, with each peak corresponding to a different photon number in a cavity~\cite{rpn,nmp}. 
Furthermore, it is known that the area ratio of the peaks obeys the photon number distribution in the cavity~\cite{qip}. 
In practice, however, we find the effect of the finite power of the qubit drive field, which gives rise to a discrepancy between the observed peak area ratio and the actual photon number distribution. 
At the same time, it turns out that the qubit drive actually enhances the signal-to-noise ratio of the photon number peaks in the qubit spectrum.
By fitting the obtained spectrum with a model which takes into account the effect, we determine the actual photon number distribution.
The photon number distribution confirms its nonclassicality by Klyshko's criterion, quantitatively indicating an even-odd photon number oscillation~\cite{kly}.
This is a steady-state realization and characterization of a nonclassical photon number distribution in a cavity which is continuously driven by a squeezed vacuum.
Owing to the input-output relation~\cite{iqn}, the photon number distribution in the cavity can be interpreted as that of the injected microwave state in a propagating mode.
It is in stark contrast with the dynamical generations and characterizations of nonclassical states (e.g., cat states) in a cavity~\cite{saq,deq}.

\begin{figure}[t]
\begin{center}
  \includegraphics[width=80mm]{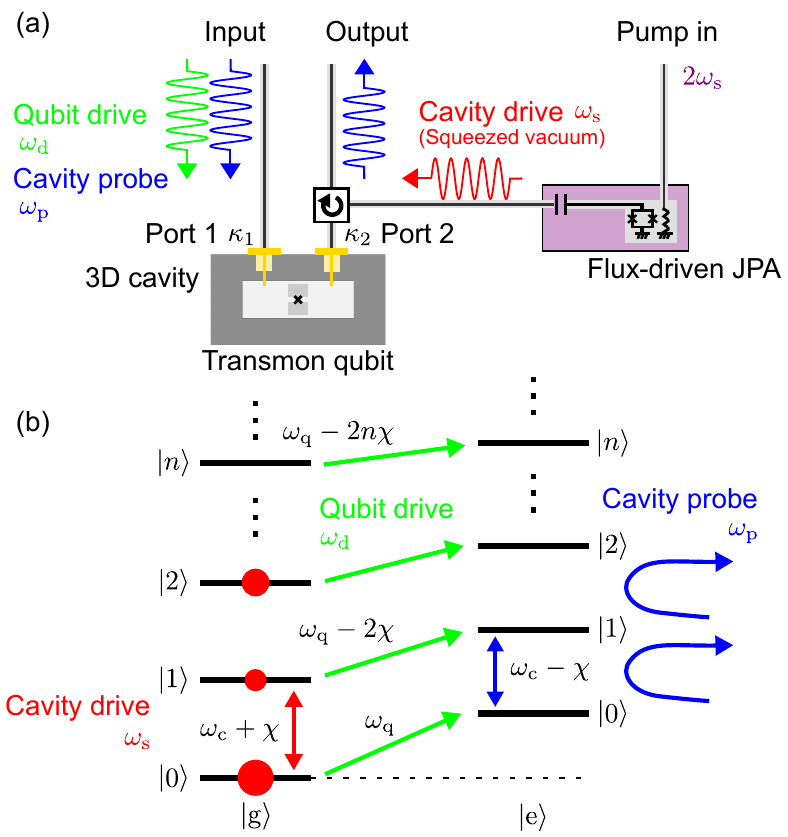} 
\caption{
(a)~Schematic of the experimental setup with squeezed vacuum injection.
A squeezed vacuum generated by a flux-driven Josephson parametric amplifier (JPA), as a cavity drive field at $\omega_{\rm s}$, is injected into the cavity from port~2.
The cavity probe field at $\omega_{\rm p}$ and the qubit drive field at $\omega_{\rm d}$ are input from port~1, and the transmission of the cavity probe field is measured. 
The cavity is designed to have asymmetric external coupling rates of $\kappa_2\approx 100\times \kappa_1$.
For the thermal- and coherent-state injections, the connection to the JPA is switched to a heavily attenuated microwave line connected to the respective sources at room temperature.
(b)~Energy levels of a dispersively coupled qubit-cavity system.
$|{\rm g}\rangle$ and $|{\rm e}\rangle$ label the ground and the first exited states of the transmon qubit,
and $|n\rangle\;(n=0,1,2,\cdots)$ indicates the photon number states of the cavity.
The cavity drive field generates the steady-state photon number distribution in the cavity (red dots).
} 
  \label{fig1} 
\end{center}
\end{figure}

We use a circuit-QED system in the strong dispersive regime, where a transmon qubit is mounted at the center of a three-dimensional superconducting cavity as shown schematically in Fig.~\ref{fig1}(a).
Setting $\hbar=1$, the qubit-cavity coupled system is described by the Hamiltonian
\begin{equation}
\label{dis}
\mathcal{H} =\omega_{\rm c} a^\dag a+\frac{\omega_{\rm q}}{2} \sigma_z-\chi a^\dag a \sigma_z,
\end{equation}
where 
$a^\dag (a)$ is the creation (annihilation) operator of the cavity mode, 
$\sigma_z$ is the Pauli operator of the transmon qubit, 
$\omega_{\rm c}/2\pi=10.4005~{\rm GHz}$ is the cavity resonant frequency, 
$\omega_{\rm q}/2\pi=8.7941~{\rm GHz}$ is the qubit resonant frequency, and
$\chi/2\pi=3.9~{\rm MHz}$  is the dispersive shift. 
Note that the Hamiltonian is truncated to the subspace of the ground state $|{\rm g}\rangle$ and the first excited state $|{\rm e}\rangle$ of the transmon qubit; the higher excited states of the qubit are not populated in the experiment below.
The total decay rate of the cavity is $\kappa/2\pi=0.5~{\rm MHz}$, 
the relaxation time of the qubit is $T_1=5.5~\mu{\rm s}$, 
and the dephasing  time of the qubit is $T^*_2=4.5~\mu{\rm s}$,
determined respectively from independent measurements.
As shown in Fig.~\ref{fig1}(b), the dispersive interaction produces both, the qubit-state-dependent shift  of the cavity resonant frequency and the photon-number-dependent light shift of the qubit resonant frequency (discrete ac Stark shift).

\begin{figure}[t]
\begin{center}
  \includegraphics[width=80mm]{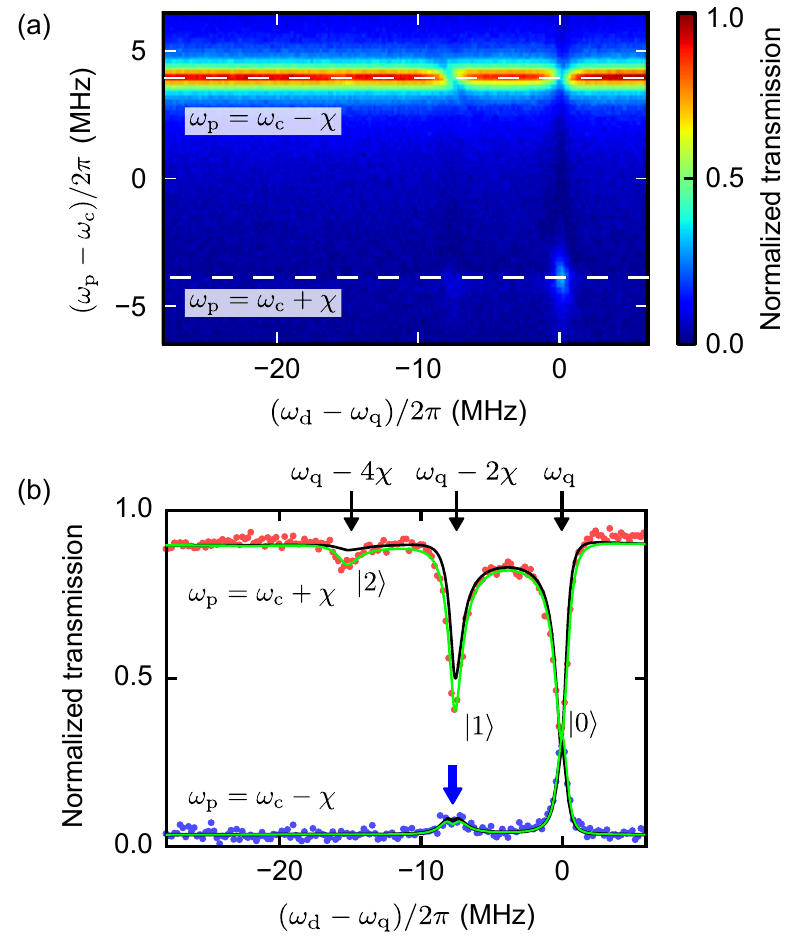} 
\caption{
(a)~Cavity transmission as a function of the qubit drive frequency $\omega_{\rm d}$ and the cavity probe frequency $\omega_{\rm p}$. The transmission is normalized by the maximum peak value. White dashed lines indicate $\omega_{\rm p}=\omega_{\rm c}\pm\chi$.
(b)~Cross sections of (a) at $\omega_{\rm p} =\omega_{\rm c} \pm \chi$ (red and blue dots, respectively). 
Green lines represent the rigorous numerical results 
in which the finite cavity probe power is fully incorporated,
whereas the black lines represent the numerical results within the linear response to the cavity probe field,
which corresponds to the weak power limit of  the cavity probe field.
The splitting of the single-photon peak, which is observed for $\omega_{\rm p} = \omega_{\rm c} - \chi$ (blue arrow), is understood as the Autler-Townes effect of the qubit, driven strongly at $\omega_{\rm d}=\omega_{\rm q}-2\chi$ (see \cite{Supple} for the details).
} 
  \label{fig2}
\end{center}
\end{figure}

\begin{figure}[t]
\begin{center}
  \includegraphics[width=80mm]{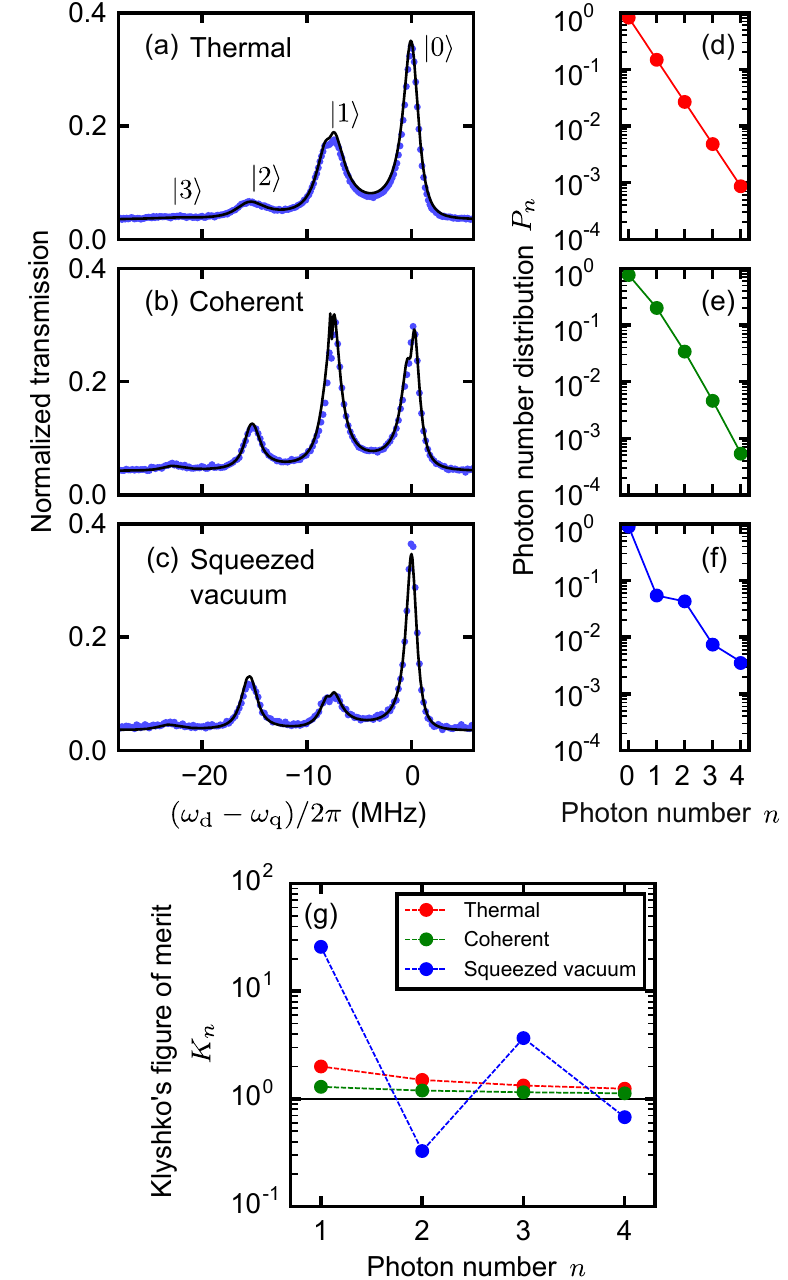} 
\caption{
(a)-(c)~Qubit spectra reflecting the photon number distributions in the cavity.
The cavity drive fields at frequency $\omega_{\rm s}$ are in (a)~thermal, (b)~coherent, and (c)~squeezed vacuum states, respectively. 
The average photon number in each state is set to about 0.2.
Blue dots are the experimental data, and the black solid lines are the numerically calculated linear responses. 
(d)-(f)~Photon number distributions determined from the fittings (dots).
Solid lines are the photon number distributions calculated from the corresponding models.
(g)~Klyshko's figure of merit $K_n$ evaluated for each drive.} 
  \label{fig3} 
\end{center}
\end{figure}

In our experiment, three inputs of continuous microwaves are used: a cavity drive, a qubit drive and a cavity probe (see Fig.~\ref{fig1}).
The cavity drive field, whose frequency $\omega_{\rm s}$ is fixed at the cavity resonant frequency for the qubit in the ground state, $\omega_{\rm c}+\chi$, is injected to the cavity to generate the steady-state photon number distribution. 
The qubit drive field is applied to the qubit whose excitation probability depends on the photon number distribution in the cavity.
The cavity probe field, whose frequency $\omega_{\rm p}$ is fixed around the cavity resonant frequency, is used to probe the transmission of the cavity depending on the the qubit excitation probability.
By measuring the cavity transmission as a function of the qubit drive frequency $\omega_{\rm d}$, we can observe a qubit spectra reflecting the photon number distribution in the cavity.
In the cavity drive field, we use a different kind of states, such as thermal states, coherent states, and squeezed vacuum states.
Thermal states are generated by amplifying the thermal noise at room temperature, 
and coherent states are generated by a microwave source at room temperature.
They are led to the cavity through a series of attenuators to suppress the background noise.
Squeezed vacuum states are generated by pumping a flux-driven JPA~\cite{fdj} at twice the JPA resonant frequency as shown in Fig.~\ref{fig1}(a).
The correlated photon pairs, generated from individual pump photons, result in an even-odd photon number oscillation in the photon number distribution.
Note that the squeezed vacuum field propagating through the waveguide has a bandwidth broader than the cavity, and the photon pairs are generated symmetrically in frequency with respect to the center frequency of the squeezed vacuum in order to conserve energy.

First of all, we study the effect of the cavity probe field on qubit spectra.
In Fig.~\ref{fig2}(a), we plot the cavity transmission as a function of the cavity probe frequency $\omega_{\rm p}$ and the qubit drive frequency $\omega_{\rm d}$.
The red (blue) dots in Fig.~\ref{fig2}(b) depict the cross-section at $\omega_{\rm p}=\omega_{\rm c}+\chi$ ($\omega_{\rm p}=\omega_{\rm c}-\chi$) in Fig.~\ref{fig2}(a).
Despite the absence of the cavity drive field at $\omega_{\rm s}$,
we observe unexpected dips and peaks corresponding to single or double photon occupation in the cavity.
Nevertheless, the numerical results obtained from the master equation taking into account the finite qubit drive and cavity probe power, reproduce these spectra very well (green lines)~\cite{Supple}.
The excess dips in the spectrum at $\omega_{\rm p}=\omega_{\rm c}+\chi$ (cavity resonant frequency for the qubit in the ground state) are induced by the back-action of the cavity probe field on the cavity transmission.
On the other hand, for $\omega_{\rm p} = \omega_{\rm c}-\chi$ (cavity resonant frequency for the qubit in the excited state), the back-action is minimal.
Note that the small single-photon peak still remains due to the thermal background noise, corresponding to the average photon number $n_{\rm th}=0.04$ in the cavity.
The black solid lines in Fig.~\ref{fig2}(b) represent the numerical results within the linear response to the cavity probe field, which corresponds to the weak power limit of the probe \cite{Supple}.
The deviation of the linear response from the observed spectrum is smaller at $\omega_{\rm p} = \omega_{\rm c} - \chi$ than at  $\omega_{\rm p} = \omega_{\rm c} + \chi$.
For the measurements below, we fix the cavity probe frequency $\omega_{\rm p} = \omega_{\rm c} - \chi$ which does not influence the qubit spectra significantly and apply the linear-response analysis.

\begin{figure}[t]
\begin{center}
  \includegraphics[width=80mm]{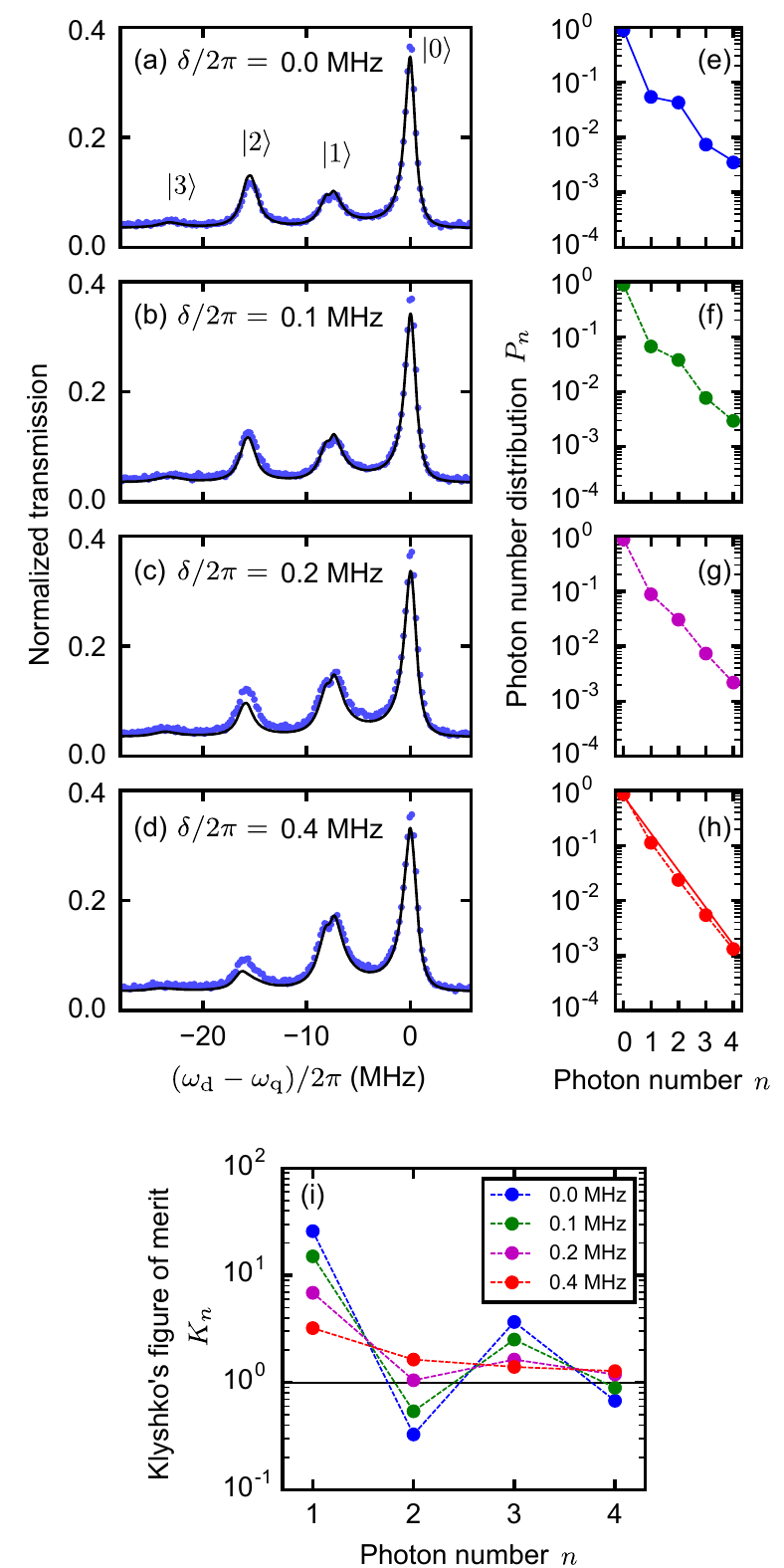} 
\caption{
(a)-(d) Squeezed-drive-frequency dependence of the qubit spectrum. 
$\delta=\omega_{\rm s}-(\omega_{\rm c}+\chi)$ is the detuning between the center frequency $\omega_{\rm s}$ of the squeezed vacuum field and the cavity resonant frequency $\omega_{\rm c}+\chi$. 
Blue dots are the experimental results, and black solid lines are the numerical calculations.
(e)-(h)~Photon number distributions determined from the fittings (dots and dashed lines).
Solid lines in (e) and (h) are the photon number distributions calculated from the corresponding models.
(i)~Klyshko's figure of merit $K_n$ evaluated for each detuning $\delta$.} 
  \label{fig4} 
\end{center}
\end{figure}

Qubit spectra obtained in the cavity driven by different states of microwave fields are shown in Figs.~\ref{fig3}(a)-(c).
The numerical calculations (black solid lines) reproduce well the experimental results (blue dots). 
Dots in Figs.~\ref{fig3}(d)-(f) represent the photon number distributions in the cavity, determined from the numerical fits for the spectra.
We compare them with the expectations based on simple models~\cite{Supple}.
The red line in Fig.~\ref{fig3}(d) is the distribution of a thermal state with the average photon number $n_{\rm th}=0.22$.
The green line in Fig.~\ref{fig3}(e) is the distribution of a thermal coherent state with $n_{\rm th} = 0.04$ and the displacement parameter $\alpha= 0.49$.
An even-odd photon number oscillation is observed both in the qubit spectrum and in the photon number distribution for the squeezed vacuum state [Figs.~\ref{fig3}(c) and (f)].
The blue line in Fig.~\ref{fig3}(f) is the distribution of a squeezed vacuum state with the squeezing parameter $r=0.54$ and the loss ratio $l=0.42$.
This corresponds to a 2.1-dB squeezed state. 
Note that the determined photon number distributions have much less weights for larger $n$ than the apparent peak area ratio in the qubit spectra.
This is because the qubit excitation rate and the cavity decay rate are larger than the qubit decay rate.
In the steady-state measurement, once the qubit is excited in the presence of the cavity photons $(n \ge 1)$, the photons leave the cavity rapidly and the population accumulates in the state $|{\rm e},0\rangle$.
Therefore, the cavity transmission signal conditioned on the qubit excited state is enhanced.

To verify the nonclassicality of the photon number distribution under the squeezed drive,
we evaluate Klyshko's figures of merit $K_n=\frac{(n+1)P_{n-1}P_{n+1}}{n{P_n}^2}$ ($n = 1,\, 2,\, \cdots$)~\cite{kly} shown in Fig.~\ref{fig3}(g).
A set of $K_n$ gives a nonclassicality criterion which can be calculated with the photon number distribution alone.
If any of $K_n$ is less than unity, the state is determined to be nonclassical.
As shown in Fig.~\ref{fig3}(g), 
$K_n$ is below unity for $n=2$ and $4$ under the squeezed drive.
Thus, the photon number distribution fulfills Klyshko's criterion for nonclassicality.
In contrast, all the values of $K_n$ up to 4 are found to be larger than unity for the coherent and the thermal drives.

Finally, we study the squeezed-drive-frequency dependence of the qubit spectrum as shown in Figs.~\ref{fig4}(a)-(d).
When the detuning $\delta$ between the center frequency $\omega_{\rm s}$ of the broadband squeezed vacuum and the cavity resonant frequency $\omega_{\rm c}+\chi$ is zero, 
both photons in a pair are injected into the cavity with a high and identical probability,
so that the even-odd photon number oscillation is conserved.
When the detuning is increased, however, 
the injection probabilities of the photon pairs are asymmetrically biased, and the photon number oscillation is weakened.
In the large detuning limit, the cavity state becomes a thermal state.
This can be understood from the fact that a two-mode squeezed vacuum state is reduced to a thermal state after tracing out one of the modes.
In Fig.~\ref{fig4}, we observe that the photon number oscillation is diminished as the detuning is increased.
Eventually, the photon number distribution approaches the Boltzmann distribution of a thermal state with the average photon number $n_{\rm th}=0.27$ [red solid line in Fig.~\ref{fig4}(h)].
These observations indicate that a broadband squeezed vacuum has correlated photon pairs in frequency space.
Klyshko's figures of merit plotted in Fig.~\ref{fig4}(i) show that the nonclassicality is reduced as the detuning is increased and that the cavity state becomes a classical state, i.e., $K_n > 1$ for any photon number $n$. 

In conclusion, we developed a circuit-QED scheme to characterize a microwave squeezed vacuum in the Fock basis.
By analyzing the qubit spectrum in a cavity driven continuously by a squeezed vacuum, we determined the photon number distribution, which is associated with the squeezed vacuum in a propagating mode according to the input-output relation.
Most importantly, the distribution fulfills Klyshko's criterion for nonclassicality.

\begin{acknowledgments} 
We acknowledge the fruitful discussion with K.\ Wakui. 
This work was supported in part by the Project for Developing Innovation System of MEXT, ALPS, JSPS KAKENHI (No.~16K05497 and 26220601), and JST ERATO (Grant No.~JPMJER1601). 
\end{acknowledgments}

\clearpage

\renewcommand{\theequation}{S\arabic{equation}}
\renewcommand{\thefigure}{S\arabic{figure}}
\renewcommand{\thetable}{S\arabic{table}}
\setcounter{equation}{0}
\setcounter{figure}{0}
\setcounter{table}{0}
\section*{Supplementary materials}

\section*{S1. Experimental setup}
\begin{figure}[t]
\bec
  \includegraphics[width=80mm]{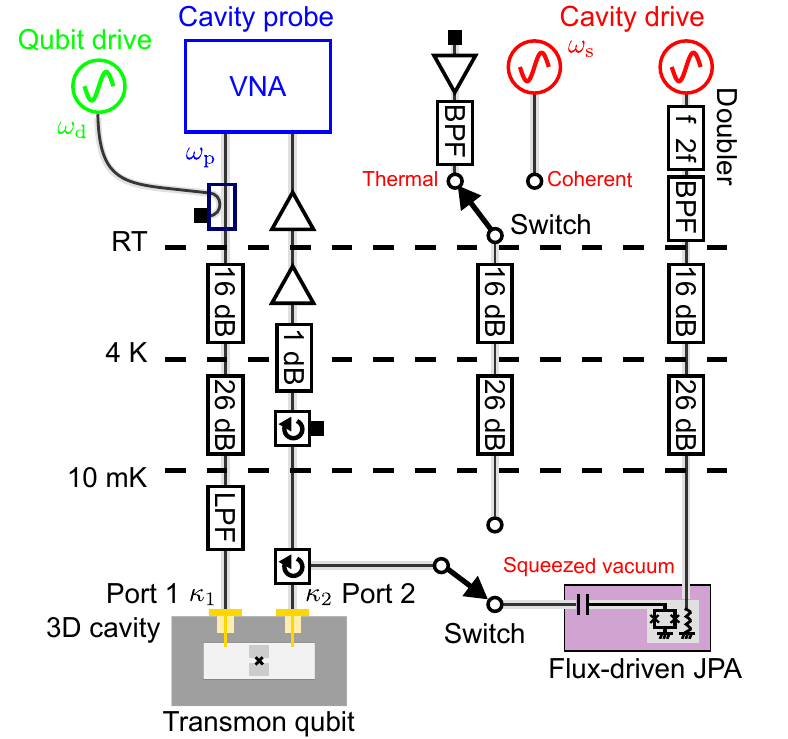} 
\caption{
Schematic of the experimental setup.
$\om_{\rm p}=\om_{\rm c}-\chi$ is the cavity probe frequency, 
$\om_{\rm d}$ is the qubit drive frequency, 
and $\om_{\rm s}=\om_{\rm c}+\chi$ is the cavity drive frequency.
}
  \label{figs1}
\enc
\end{figure}

The experimental setup is shown in Fig.~\ref{figs1}.
We use a circuit-QED architecture, where a transmon qubit is mounted at the center of a three-dimensional superconducting cavity.
The three-dimensional superconducting cavity is made of aluminum (A1050).
The transmon qubit with an $\rm{Al/AlO_{\it x}/Al}$ Josephson junction is fabricated on a silicon substrate. 
From the frequency-domain measurements (see Sec.~S4 below),
we determine
the dressed cavity resonant frequency $\omega_{\rm c}/2\pi=10.4005~{\rm GHz}$, 
the total cavity decay rate $\kap/2\pi=0.494~{\rm MHz}$,
and the effective dispersive shift $\chi/2\pi=3.9~{\rm MHz}$.
The dressed qubit resonant frequency is $\omega_{\rm q}/2\pi=8.7941~{\rm GHz}$,
and the dressed anharmonicity is $-136~{\rm MHz}$. 
By using the dressed frequencies,
we find the bare cavity resonant frequency, $10.3660~{\rm GHz}$, and the coupling strength between the qubit and the cavity, $240~{\rm MHz}$.
The bare qubit resonant frequency is $8.8320~{\rm GHz}$, 
and its anharmonicity is $-140~{\rm MHz}$, corresponding to $E_{\rm J}/E_{\rm C}\approx 500$.

To determine the photon number distribution in the cavity that is driven continuously by various types of microwave fields, we observe the qubit spectrum.
The transmission of the cavity probe field at frequency $\om_{\rm p}$ is measured by using a vector network analyzer (VNA), while sweeping the qubit drive frequency at frequency $\om_{\rm d}$.
The qubit drive field is added to the input line at a directional coupler at room temperature.
We use thermal states, coherent states, and squeezed vacuum states for the cavity drive field. 
A switch connects each source to the cavity, as shown in Fig.~\ref{figs1}. 
Thermal states are generated by amplifying and filtering the thermal noise at room temperature and led to the cavity through attenuators.
Coherent states are generated by a microwave source at room temperature and led to the cavity through attenuators to suppress the thermal background.
Squeezed vacuum states are generated by pumping a flux-driven Josephson parametric amplifier (JPA) at twice the JPA resonant frequency \cite{fdj,sfj}.
The JPA resonant frequency is tuned to the cavity resonant frequency with the qubit in the ground state by applying the DC magnetic field.

\section*{S2. Theoretical description}
\begin{figure}[t]
\bec
\includegraphics[width=80mm]{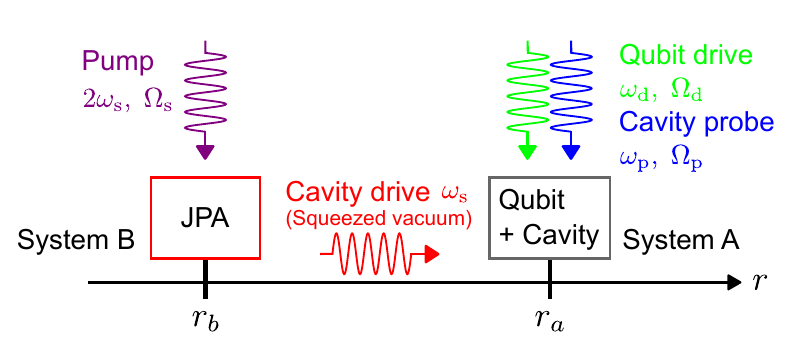}
\caption{
Schematic of the model. 
The output field from System~B (JPA) is used as the input for System~A (qubit and cavity).
The coordinate $r$ is defined along the propagation direction of the waveguide field.
A circulator placed between Systems A and B in the experiment (see Fig.~S1) allows us to treat only the right-going mode in the calculation.
}
\label{figs2}
\enc
\end{figure}

In this section, we present the formula used in the numerical calculations.
In the setup considered (Fig.~\ref{figs2}), 
a qubit-cavity system (System~A) is subject to a continuous squeezed vacuum field generated by a JPA (System~B).
Setting $\hbar=1$, The Hamiltonian of System~A with the qubit drive and the cavity probe fields is described by
\beq
\label{h}
\begin{split}
\cH =&\om_{\rm c}a^\dag a +\frac{\om_{\rm q}}{2}\s_z-\chi a^\dag a\:\s_z
+ \frac{\Om_{\rm d}}{2}(e^{-i\om_{\rm d} t}\s^{\dag} + e^{i\om_{\rm d} t}\s)\\
&+ \frac{\Om_{\rm p}}{2}(e^{-i\om_{\rm p} t}a^{\dag} + e^{i\om_{\rm p} t}a),
\end{split}
\eeq
where
$a$ and $\sigma$ respectively denote the annihilation operators of the cavity mode and the qubit, 
$\sigma_z=\sigma^{\dagger}\sigma-\sigma\sigma^{\dagger}$,
and $\Om_{\rm d}$ and $\Om_{\rm p}$ are the amplitudes of the qubit drive and the cavity probe, respectively.
The Hamiltonian of System~B is given by
\beq
\cH' = \om_{\rm s} b^{\dag}b
+ \frac{\Om_{\rm s}}{2}(e^{-2i\om_{\rm s} t}b^{\dag 2} + e^{2i\om_{\rm s} t}b^2), 
\label{HB}
\eeq
where $b$ is the annihilation operator of the JPA mode, 
and $\om_{\rm s}$ is its frequency.
We apply a pump field with frequency $2\om_{\rm s}$ and amplitude $\Om_{\rm s}$ to the JPA 
to generate a squeezed vacuum. 

By taking the free Hamiltonian $\cH_0 = \om_{\rm s}(a^{\dag}a + b^{\dag}b) + \frac{\om_{\rm d}}{2} \s_z$, 
we switch to the rotating frame. 
In this frame, $\cH$ and $\cH'$ are rewritten as
\bea
\cH &=& 
(\om_{\rm c}-\om_{\rm s})a^\dag a +\frac{(\om_{\rm q}-\om_{\rm d})}{2}\s_z-\chi a^\dag a\s_z
\label{HA2} \nonumber \\
&+& \frac{\Om_{\rm d}}{2}(\s^\dag +\s)+\frac{\Om_{\rm p}}{2}(e^{-i(\om_{\rm p}-\om_{\rm s}) t}a^{\dag} + e^{i(\om_{\rm p}-\om_{\rm s})t}a),
\nonumber \\
& &\\
\cH' &=& \frac{\Om_{\rm s}}{2}(b^{\dag 2} + b^2).
\label{HB2}
\eea

The squeezed vacuum field generated by System~B is guided to System~A through a waveguide.
We define a spatial coordinate $r$ along with the propagating direction 
of the waveguide field (see Fig.~\ref{figs2}).
The waveguide field interacts with System~A at $r_a$ and System~B at $r_b$. 
Setting the microwave velocity in the waveguide to unity,
the overall Hamiltonian is written as 
\bea
\cH_{\rm total} &=& \cH + \cH' + \int dk \ kc_k^{\dag}c_k \label{total} \nonumber \\
&+& \sqrt{\kap'_{\rm e}}\left( b^{\dag}\tc_{r_b} + \tc^{\dag}_{r_b}b \right)
+ \sqrt{\kap_{\rm e}}\left( a^{\dag}\tc_{r_a} + \tc^{\dag}_{r_a}a \right), \nonumber \\
& &
\eea
where
$c_k$ is the waveguide-field operator with wave number $k$,
and $\kap_{\rm e}$ ($\kap'_{\rm e}$) represents the external coupling of System~A (B) to the waveguide field.
$\tc_r$ is the spatial representation of the waveguide-field operator,
as given by $\tc_r=(2\pi)^{-1/2}\int dk e^{ikr}c_k$. 
Note that the photon frequency should be measured relative to $\om_{\rm s}$, 
since we are in the rotating frame.

We denote an arbitrary operator belonging to System~A (B) by $S_{\rm A}$ ($S_{\rm B}$)
and investigate its time evolution at $t$ ($t-l$), 
where $l=r_a-r_b(>0)$ is the distance between the two systems.  
This is because $S_{\rm A}(t)$ and $S_{\rm B}(t-l)$ are on the same light cone
and are therefore relativistically simultaneous. 
From Eq.~(\ref{total}), we can derive the following Heisenberg equations,
\bea
\frac{d}{dt}S_{\rm A} = i[\cH, S_{\rm A}] 
&+& i\sqrt{\kap_{\rm e}}\:[a^{\dag},S_{\rm A}]\tc_{r_a}(t) \nonumber \\ 
&+& i\sqrt{\kap_{\rm e}}\:\tc_{r_a}^{\dag}(t)[a,S_{\rm A}], 
\label{dsa} \\
\nonumber \\
\frac{d}{dt}S_{\rm B} = i[\cH', S_{\rm B}] 
&+& i\sqrt{\kap'_{\rm e}}\:[b^{\dag},S_{\rm B}]\tc_{r_b}(t-l)  \nonumber \\
&+& i\sqrt{\kap'_{\rm e}}\:\tc_{r_b}^{\dag}(t-l)[b,S_{\rm B}], 
\label{dsb}
\eea
and the input-output relation, 
\bea
\tc_r(t) = &\tc_{r-t}(0)&  
\label{io} \nonumber \\
&-&i\sqrt{\kap'_{\rm e}}\:\theta(r-r_b)\theta(t-r+r_b)b(t-r+r_b) \nonumber \\ 
&-&i\sqrt{\kap_{\rm e}}\:\theta(r-r_a)\theta(t-r+r_a)a(t-r+r_a), \nonumber \\
& & 
\eea
where 
$\theta(t)$ is the step function. 
Since we analyze the stationary response, we assume that $t$ is sufficiently large. 
Therefore,
$\tc_{r_b}(t-l)=\tc_{r_a-t}(0)-i\sqrt{\kap'_{\rm e}}\:b(t-l)/2$ and 
$\tc_{r_a}(t)=\tc_{r_a-t}(0)-i\sqrt{\kap_{\rm e}}\:a(t)/2-i\sqrt{\kap'_{\rm e}}\:b(t-l)$. 
From these equations, Eqs.~(\ref{dsa}) and (\ref{dsb}) are rewritten as
\bea
\frac{d}{dt}S_{\rm A}&=& i[\cH, S_{\rm A}] 
+ \frac{\kap_{\rm e}}{2}\cL_a[S_{\rm A}] 
\label{dsa2} \nonumber \\
&+& \sqrt{\kap_{\rm e}\kap'_{\rm e}}\:[a^{\dag},S_{\rm A}]b + \sqrt{\kap_{\rm e}\kap'_{\rm e}}\:b^{\dag}[S_{\rm A},a]  \nonumber \\ 
&+& i\sqrt{\kap_{\rm e}}\:[a^{\dag},S_{\rm A}]\tc_{r_a-t}(0) + i\sqrt{\kap_{\rm e}}\:\tc^{\dag}_{r_a-t}(0)[a,S_{\rm A}],  \nonumber \\
& & \\
\frac{d}{dt}S_{\rm B} &=& i[\cH', S_{\rm B}] 
+ \frac{\kap'_{\rm e}}{2}\cL_b[S_{\rm B}] 
\label{dsb2} \nonumber \\
&+& i\sqrt{\kap'_{\rm e}}\:[b^{\dag},S_{\rm B}]\tc_{r_a-t}(0)+ i\sqrt{\kap'_{\rm e}}\:\tc^{\dag}_{r_a-t}(0)[b,S_{\rm B}], \nonumber \\
& &
\eea
where 
$\cL_a[S_{\rm A}]=[a^{\dag},S_{\rm A}]a+a^{\dag}[S_{\rm A},a]$. 
The Heisenberg equation for the product operator $S_{\rm B}S_{\rm A}$ can be derived from Eqs.~(\ref{dsa2}) and (\ref{dsb2}). 
Care should be taken that 
$[\tc_{r_a-t}(0), S_{\rm B}(t-l)] = i\sqrt{\kap'_{\rm e}}\:[b(t-l),S_{\rm B}(t-l)]/2$ and 
$[\tc_{r_a-t}(0), S_{\rm A}(t)] = i\sqrt{\kap_{\rm e}}\:[a(t),S_{\rm A}(t)]/2$, 
both of which result from Eq.~(\ref{io}). 
In the considered setup, 
we do not apply an input field to System~A through the waveguide. 
Therefore, 
denoting the initial state vector of the overall system by $|\psi_{\rm i}\ra$, 
we can rigorously take $\tc_{r}(0)|\psi_{\rm i}\ra=0$. 
Then, the equation of motion for  
$\la S_{\rm A}S_{\rm B} \ra=\la \psi_{\rm i}|S_{\rm A}S_{\rm B}|\psi_{\rm i}\ra$ 
is written as
\bea
\frac{d}{dt}\la S_{\rm A}S_{\rm B} \ra &=& 
i\la[\cH, S_{\rm A}]S_{\rm B}\ra + i\la S_{\rm A}[\cH', S_{\rm B}]\ra 
\label{mean} \nonumber \\
&+&\sqrt{\kap_{\rm e}\kap'_{\rm e}}\:\la [S_{\rm A},a]b^{\dag}S_{\rm B} \ra
+\sqrt{\kap_{\rm e}\kap'_{\rm e}}\:\la [a^{\dag},S_{\rm A}]S_{\rm B}b \ra 
\nonumber \\
&+& \frac{\kap_{\rm e}}{2}\la \cL_a[S_{\rm A}] S_{\rm B} \ra
+ \frac{\kap'_{\rm e}}{2}\la S_{\rm A}\cL_b[S_{\rm B}] \ra. 
\eea

Up to here, we assumed for simplicity that Systems~A and B damp 
only through the radiative coupling to the waveguide field.
Here, we include other dissipation channels, such as 
the decay of cavities~A and B into other environments,  
and the decay and pure dephasing of the qubit in System~A. 
Furthermore, we take account of the thermal excitation of the systems through the environment.
Then, Eq.~(\ref{mean}) should be replaced with the following one, 
\bea
\frac{d}{dt}\la S_{\rm A}S_{\rm B} \ra &=& 
  i\la[\cH, S_{\rm A}]S_{\rm B}\ra
+i\la S_{\rm A}[\cH', S_{\rm B}]\ra 
\label{fin} \nonumber \\
&+&  \sqrt{\kap_{\rm e}\kap'_{\rm e}}\: \la [S_{\rm A},a]b^{\dag}S_{\rm B}  \ra
+  \sqrt{\kap_{\rm e}\kap'_{\rm e}} \:\la [a^{\dag},S_{\rm A}]S_{\rm B} b \ra \nonumber \\
&+& \frac{\kap'_{\rm e}}{2}\la S_{\rm A}\cL_b[S_{\rm B}] \ra 
+ \frac{\kap (1+n_{\rm th})}{2}\la \cL_a[S_{\rm A}] S_{\rm B} \ra \nonumber \\
&+& \frac{\kap n_{\rm th}}{2}\la \cL_{a^{\dag}}[S_{\rm A}] S_{\rm B} \ra
+ \frac{\gam(1+p_{\rm th})}{2}\la \cL_{\s}[S_{\rm A}] S_{\rm B} \ra \nonumber \\
&+& \frac{\gam p_{\rm th}}{2}\la \cL_{\s^{\dag}}[S_{\rm A}] S_{\rm B} \ra 
+ \frac{\gam_\phi}{2}\la \cL_{\s^{\dag}\s}[S_{\rm A}] S_{\rm B} \ra. \nonumber \\
& &
\eea
where
$\kappa$ is the total cavity decay, 
$n_{\rm th}$ is the average thermal photon number in the cavity,
$\gam=1/T_1$ is the qubit decay rate,
$p_{\rm th}$ is the thermal excitation probability of the qubit,
and $\gam_{\phi}$ is the qubit pure dephasing rate.
Note that the internal loss and the thermal photon excitation of the JPA mode are neglected in Eq.~(\ref{fin}).

In the Fock-state basis, 
the state vector of the composite system is written as $|q,n,m\rangle$, 
where $q(={\rm g,e})$ denotes the qubit state in System~A,
and $n$ and $m(=0,1,\cdots)$ denotes 
the cavity photon numbers in Systems~A and B, respectively. 
The density matrix of the composite system is obtained by
setting $S_{\rm A}S_{\rm B}=S_{qnm,q'n'm'}=|q,n,m\rangle\langle q',n',m'|$
in Eq.~(\ref{fin}). 
Since the probe field is weak, 
we solve this equation perturbatively in $\Om_{\rm p}$. 
For this purpose, we first determine the steady-state solution 
$\la S_{qnm,q'n'm'} \ra^{(0)}$ by setting $\Om_{\rm p}=0$ in Eq.~(\ref{fin}). 
Then, we determine the linear response $\la S_{qnm,q'n'm'} \ra^{(1)}$, 
which is proportional to $\Om_{\rm p} e^{-i(\om_{\rm p}-\om_{\rm d})t}$. 
Since the output probe field is measured at a different port (Port~2 in Fig.~\ref{figs1}) from the input one (Port~1 in Fig.~\ref{figs1}), 
the probe transmission coefficient is proportional to
the cavity amplitude of System~A,
$\la a \ra^{(1)}=\sum_{q,n,m}\sqrt{n+1}\la S_{qnm,q(n+1)m} \ra^{(1)}$.
The parameters used in the numerical calculations are shown in Table~\ref{table1}.

\begin{table}[bp]
\bec
\caption{System parameters.}
  \begin{tabular}{lcl} \hline \hline
Dressed cavity resonant frequency & $\om_{\rm c}/2\pi$  & 10.4005~GHz \\ \hline
Cavity external coupling rate    & $\kap_{\rm e}/2\pi$  & 0.490~MHz \\ \hline
Cavity total decay rate                  & $\kap/2\pi$            & 0.494~MHz \\ \hline
Thermal average photon number     & $n_{\rm th}$                 & 0.04 \\ \hline
Cavity probe amplitude                  & $\Om_{\rm p}/2\pi$ & 0.16~MHz \\ \hline
Dressed qubit resonant frequency  & $\om_{\rm q}/2\pi$  & 8.7941~GHz \\ \hline 
Qubit decay rate                           & $\gam = 1/T_1$       & 1/5.5~${\mu {\rm s}}^{-1}$ \\ \hline
Qubit dephasing rate                     & $\gam_\phi$             & 0  \\ \hline
Thermal excitation probability         & $p_{\rm th}$                & 0.01 \\ \hline
Qubit drive amplitude                    & $\Om_{\rm d}/2\pi$  & 0.46~MHz \\ \hline
Effective dispersive shift               & $\chi/2\pi$                   & 3.9~MHz \\ \hline
JPA external (total) coupling rate    & $\kap'_{\rm e}/2\pi$ & 40~MHz \\ \hline \hline  
  \end{tabular}\
\label{table1}
\enc  
\end{table}
The parameters characterizing the cavity drive fields (thermal, coherent, and squeezed vacuum states) are determined by fitting the qubit spectrum with numerical results from Eq.~(\ref{fin}).
Then, the photon number distribution is determined from Eq.~(\ref{fin}) in absence of the qubit drive and cavity probe fields ($\Om_{\rm d}=\Om_{\rm p}=0$).
When a thermal state is applied as the cavity drive field,
the average thermal photon number $n_{\rm th}$ is used as the fitting parameter.
When a coherent state is used as the cavity drive,
the Hamiltonian of System~B is replaced with
\beq
\cH' = \frac{\Om_{\rm s}}{2}(b^\dag + b).
\label{HB3}
\eeq
Then, the output field from System~B becomes a coherent state.
The amplitude $\Om_{\rm s}$, corresponding to the strength of the coherent drive to the cavity, is used as an additional fitting parameter.
For the case with a squeezed vacuum drive,
we need to incorporate the loss of waveguide between the JPA and the cavity,
since the squeezed vacuum state is degraded considerably by the loss of the waveguide.
Theoretically, such waveguide loss is taken into account
by decreasing the coupling $\kappa_{\rm e}$ between the waveguide and the cavity of System~A while keeping its total decay rate $\kappa$.
Accordingly, the pump amplitude for the JPA, $\Om_{\rm s}$, and the external coupling rate of the cavity, $\kap_{\rm e}$, are used as the fitting parameters.
In the numerical simulations in Fig.~3 of the main text, we employed the following parameters: $n_{\rm th}=0.22$ in Fig.~3(a),
$\Om_{\rm s}/2\pi=1.3~{\rm MHz}$ in Fig.~3(b), and
$\Om_{\rm s}/2\pi=4.0~{\rm MHz}$ and $\kap_{\rm e}/2\pi=0.42~{\rm MHz}$ in Fig.~3(c).

\begin{figure}[t]
\bec
  \includegraphics[width=80mm]{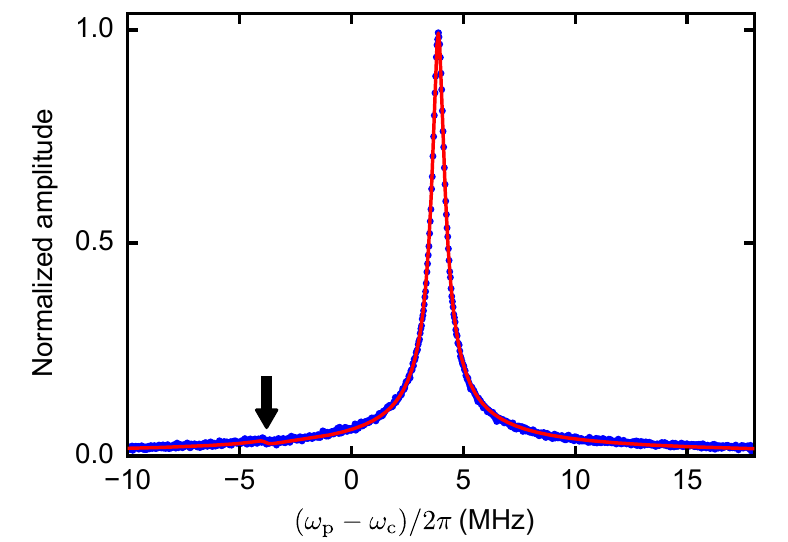} 
\caption{Cavity transmission as a function of the cavity probe frequency $\om_{\rm p}$.
The amplitude is normalized by the maximum peak value.
This normalization factor is used commonly all through the paper.
The main peak at $\om_{\rm p}=\om_{\rm c}+\chi$ corresponds to the cavity resonance with the qubit in the ground state. 
The small cavity peak at $\om_{\rm p}=\om_{\rm c}-\chi$ (arrow), corresponding to the qubit excited state, is also observed due to the thermal excitation of the qubit. 
Red solid line represents the numerical result.
} 
  \label{figs3}
\enc
\end{figure}

\section*{S3. Photon number distribution}
Throughout this work, we determine the cavity photon number distribution numerically, based on the framework described in Sec.~2. 
In order to characterize these quantum states more intuitively, we here employ the single-mode density matrices and evaluate their photon number distributions. 
When the classical fields, such as thermal and coherent states, are applied to the cavity,
the cavity state is described by a thermal coherent state, whose density matrix $\rho_{\rm tc}$ is given by
\beq
\rho_{\rm tc}= \mathcal{D}(\alpha)\rho(n_{\rm th})\mathcal{D}^\dag(\alpha),
\label{rtc}
\eeq
where $\mathcal{D}(\alpha)=\exp(\alpha a^\dag-\alpha^* a)$ is the displacement operator with a parameter $\alpha$,
and $\rho(n_{\rm th})\:\propto \left(\frac{n_{\rm th}}{1+n_{\rm th}}\right)^{a^\dag a}$ is the thermal-state density matrix with the average photon number $n_{\rm th}$.
In Figs.~3(a) and (b) of the main text, 
we plot the photon number distribution calculated from Eq.~(\ref{rtc}) by solid lines.
We find that the cavity state for the thermal drive corresponds to a thermal coherent state with $|\alpha|=0.0$ and $n_{\rm th}=0.22$, which is an exact thermal state.
In the same way, the cavity state for the coherent cavity drive is fitted by a thermal coherent state with $|\alpha|=0.49$ and $n_{\rm th}=0.04$.
The finite thermal photon population is due to the background noise from room temperature.
For the squeezed cavity drive, 
we assume the following density matrix,
\beq
\rho_{\rm sq}= {\rm Tr}_{a'}\left[\mathcal{U}_{\rm BS}(\theta)\:\rho(r)\otimes\rho'_0\:\mathcal{U}_{\rm BS}^\dag(\theta)\right],
\label{rsq}
\eeq
where $\rho(r)=\mathcal{S}(r)|0\ra\la0|\mathcal{S}^\dag(r)\;\left[\mathcal{S}(r)=\exp\left(\frac{r}{2}(a^2-a^{\dag2})\right)\right]$ is a squeezed vacuum state of the cavity mode $a$ with a squeezing parameter $r$,
$\rho_0'$ is a vacuum state of an ancilla mode $a'$,
$\mathcal{U}_{\rm BS}(\theta)=\exp\left(-\frac{\theta}{2}\left(aa'^{\dag}+a^\dag a'\right)\right)$ is a unitary operator describing a beam splitter with a loss rate of $l=\sin \theta$,
and ${\rm Tr}_{a'}$ is a partial trace for the ancilla mode $a'$.
By fitting the photon number distribution with this theoretical model,
we find the cavity quantum state with the squeezed drive corresponds to a squeezed vacuum state ($r=0.54$) with a loss ($l=0.42$).

\begin{figure}[t]
\bec
  \includegraphics[width=80mm]{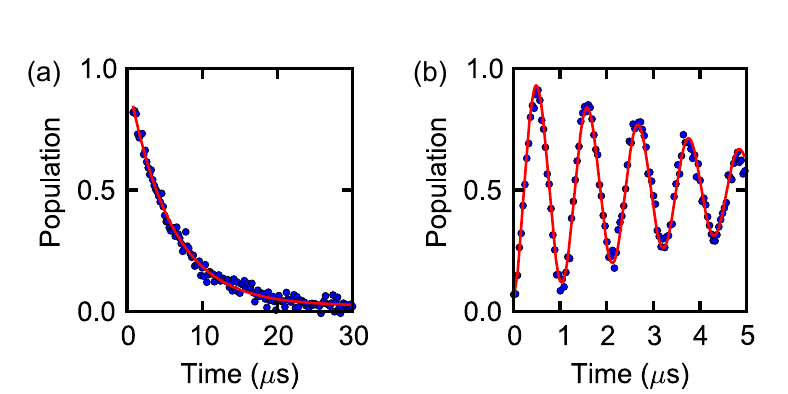} 
\caption{Time-domain measurements of the qubit coherence.
 (a)~Relaxation of the qubit.
Red solid line is a fit to a exponential curve with $T_1 = 5.5~{\mu s}$.
(b)~Dephasing of the qubit.
$T_2^*$ is $4.5~{\mu s}$.
Red solid line is the numerical result with $n_{\rm th}=0.04$.
} 
  \label{figs4}
\enc
\end{figure}

\section*{S4. System parameters}
In this section, we explain how we determined the parameters used in the previous sections.
Since the cavity drive field $\om_{\rm s}$ is absent here, we take a rotating frame determined by
$\cH_0 = \om_{\rm p} a^{\dag}a + \frac{\om_{\rm d}}{2} \s_z$.
Then, the Hamiltonian Eq.~(\ref{h}) is rewritten as 
\beq
\label{hr}
\begin{split}
\cH=(\om_{\rm c}-\om_{\rm p})a^\dag a&+\frac{(\om_{\rm q}-\om_{\rm d})}{2}\s_z-\chi a^\dag a \:\s_z\\
&+\frac{\Om_{\rm d}}{2}(\s^\dag +\s)+\frac{\Om_{\rm p}}{2}(a^\dag+a).
\end{split}
\eeq
By using Eq.~(\ref{fin}) with $S_{\rm B}=\hat{1}$ and $\kappa'_{\rm e}=0$, we calculate the time-evolutions and the steady-state solutions of System~A.
They correspond to the solution of the conventional system-bath master equation of System~A.

First, the cavity transmission amplitude, measured as a function of the cavity probe frequency $\om_{\rm p}$ in the absence of any drive field, is shown in Fig.~\ref{figs3}.
The cavity resonance is observed at the probe power corresponding to the single photon level.
The main peak at $\om_{\rm p}=\om_{\rm c}+\chi$ is the cavity resonance with the qubit in the ground state.
In addition, the small peak corresponding to the cavity with the qubit excited state is also observed at $\om_{\rm p}=\om_{\rm c}-\chi$ due to the finite thermal excitation probability of the qubit, $p_{\rm th}$.
The red solid line is calculated from the steady-state solution of Eq.~(\ref{fin}), by setting $S_{\rm A}=a$.
From this, we find $p_{\rm th}=0.01$.

Next, time-domain measurements are conducted to evaluate the coherence of the qubit.
A DAC-ADC system, instead of the VNA in Fig.~\ref{figs1}, is used for the measurement.  
The results of the qubit relaxation and Ramsey decay measurements are shown in Figs.~\ref{figs4}(a) and (b).
We obtain $T_1 = 5.5~\mu {\rm s}$, $T_2^*=4.5~\mu{\rm s}$ by fitting the data.
The total dephasing rate of the qubit $1/T_2^*$ is described with $\gam/2 + \gam_\phi + \gam_{\rm th}$,
where $\gam_{\rm th}=\frac{4\kap\chi^2}{\kap^2+\chi^2}\:n_{\rm th}$ is the dephasing rate due to the thermal photon fluctuation in the cavity \cite{qth}.
Assuming $\gam_\phi=0$,
the thermal average photon number $n_{\rm th}$ in the cavity is determined to be $0.04$ by using the simple formula.
The red solid line in Fig.~\ref{figs3}(b) is the time-evolution solution of Eq.~(\ref{fin}), by setting $S_{\rm A}=(\sigma_z+1)/2$, 
where $(\om_{\rm q}-\om_{\rm d})/2\pi=0.9~{\rm MHz}$,
$\Om_{\rm d}=\Om_{\rm p}=0$,
and $n_{\rm th}=0.04$ are used.
The calculation reproduces well the experimental result.

\begin{figure}[t]
\bec
  \includegraphics[width=80mm]{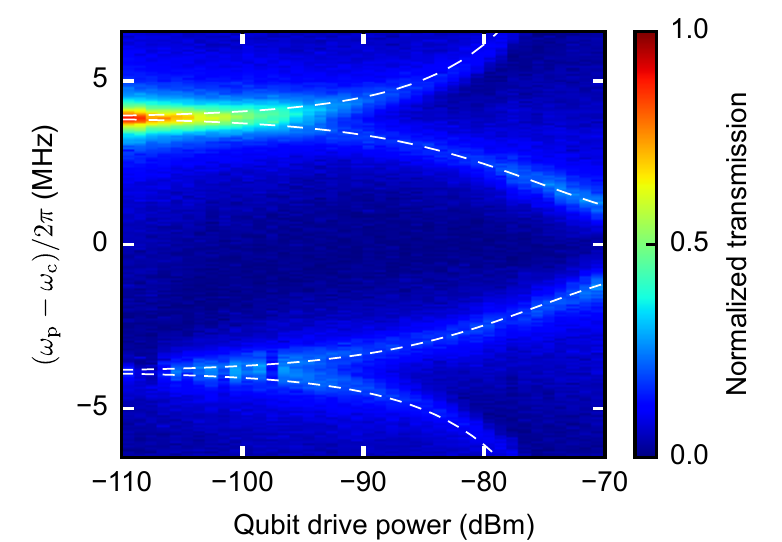} 
\caption{Cavity transmission as a function of the cavity probe frequency $\om_{\rm p}$ and the qubit drive power ($\Om_{\rm d}$).
White dashed lines depict the observed cavity resonances.
} 
  \label{figs5}
\enc
\end{figure}

\begin{figure}[t]
\bec
  \includegraphics[width=80mm]{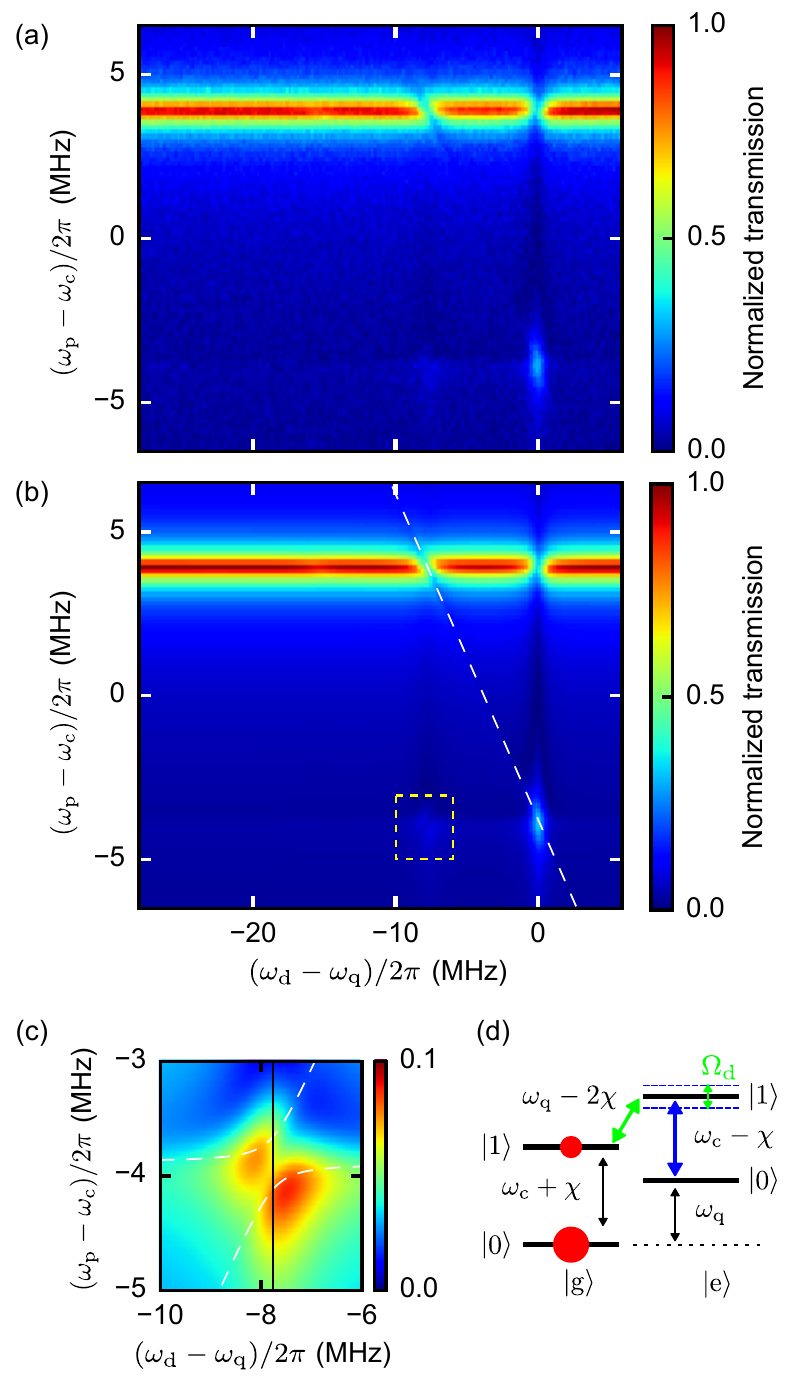} 
\caption{Cavity transmission as a function of the qubit drive frequency $\om_{\rm d}$ and the cavity probe frequency $\om_{\rm p}$.
(a)~Experimental data.
(b)~Steady-state solutions of Eq.~(\ref{fin}).
Diagonal dashed line corresponds to the resonant condition for the two-photon transition, $\omega_{\rm p} + \omega_{\rm d} =  \omega_{\rm c}+ \omega_{\rm q} -\chi$.
(c)~Magnified plot of the region in yellow dashed rectangle in (b) at around $\omega_{\rm p}= \omega_{\rm c}-\chi$ and $\omega_{\rm d} =  \omega_{\rm q} -2\chi$.
White dashed lines depict the calculated transition frequencies between $|{\rm e},0\rangle$ and the hybridized states composed of $|{\rm e},1\rangle$ and $|{\rm g},1\rangle$.
Black solid line indicates $\omega_{\rm d}=\omega_{\rm q}-2\chi$, corresponding to the qubit transition frequency with the single photon state in the cavity.
(d)~Energy levels of the dispersively coupled qubit-cavity system with the qubit drive field.
$|{\rm e},1\rangle$ and $|{\rm g},1\rangle$ are hybritized by the qubit drive.
} 
  \label{figs6}
\enc
\end{figure}

In order to calibrate the qubit drive power, the cavity transmission is measured as a function of the cavity probe frequency $\om_{\rm p}$ and the qubit drive power ($\Om_{\rm d}$), which is shown in Fig.~\ref{figs5}.
The qubit drive frequency $\om_{\rm d}$ is in resonance with the qubit resonant frequency $\om_{\rm q}$.
As the qubit drive power increases, the cavity peak corresponding to the qubit in the excited state appears, and each peak splits into two peaks due to Rabi splitting of the qubit.
In this experiment, the cavity probe power is weak enough to excite at most the single photon state in the cavity.
Therefore, the four resonances in Fig~\ref{figs5} correspond to the transitions between the lowest eigen frequencies: $\om_{0\:\pm}=\pm\frac{\Om_{\rm d}}{2}$ and $\om_{1\:\pm}=\om_{\rm c}\pm\sqrt{\chi^2+\left(\frac{\Om_{\rm d}}{2}\right)^2}$,
which are calculated from the Hamiltonian Eq.~(\ref{hr}) with $\om_{\rm d}=\om_{\rm q}$ and $\om_{\rm p}=\Om_{\rm p}=0$.
The white dashed lines in Fig.~\ref{figs5} depict these transition frequencies and agree with the observed resonance peaks.
With this plot, we performed the calibration between the actual qubit drive power and the quabit drive amplitude $\Omega_{\rm d}$.
The qubit drive power $-97~{\rm dBm}$ at sample, that we use for the measurement of the qubit spectroscopy, corresponds to  $\Om_{\rm d}/2\pi = 0.46~{\rm MHz}$. 

In order to calibrate the cavity probe power, we use the qubit spectra at $\om_{\rm p}= \om_{\rm c}\pm \chi$, as shown in Fig.~2(b) of the main text.
The red (blue) dots plot the cavity transmission as a function of the qubit drive frequency $\om_{\rm d}$, fixing the cavity probe frequency $\om_{\rm p}=\om_{\rm c}+\chi\;(\omega_{\rm p}= \om_{\rm c}-\chi)$.
The qubit spectra strongly reflect the cavity probe power $\Om_{\rm p}$ and the average thermal photon number $n_{\rm th}$ in the cavity.
The green solid lines are the steady-state solutions of Eq.~(\ref{fin}), by setting $S_{\rm A}=a$.
From the simulations, we find the cavity probe power $-125~{\rm dBm}$ at sample, that we use for the qubit spectroscopy, corresponds to  $\Om_{\rm p}/2\pi = 0.16~{\rm MHz}$.
The average thermal photon number $n_{\rm th}$, which is determined from $T_2^*$ measurement, agrees well with the qubit spectra.
 
Using these parameters, listed in Table~\ref{table1}, the cavity transmission as a function of the qubit drive frequency $\om_{\rm d}$ and the cavity probe frequency $\om_{\rm p}$ are numerically calculated from the steady-state solution of Eq.~(\ref{fin}), by setting $S_{\rm A}=a$, as shown in Fig.~\ref{figs6}.
The calculation results agree well with the experimental results, which assures the accuracy in the determination of the parameters.

\begin{figure}[t]
\bec
  \includegraphics[width=80mm]{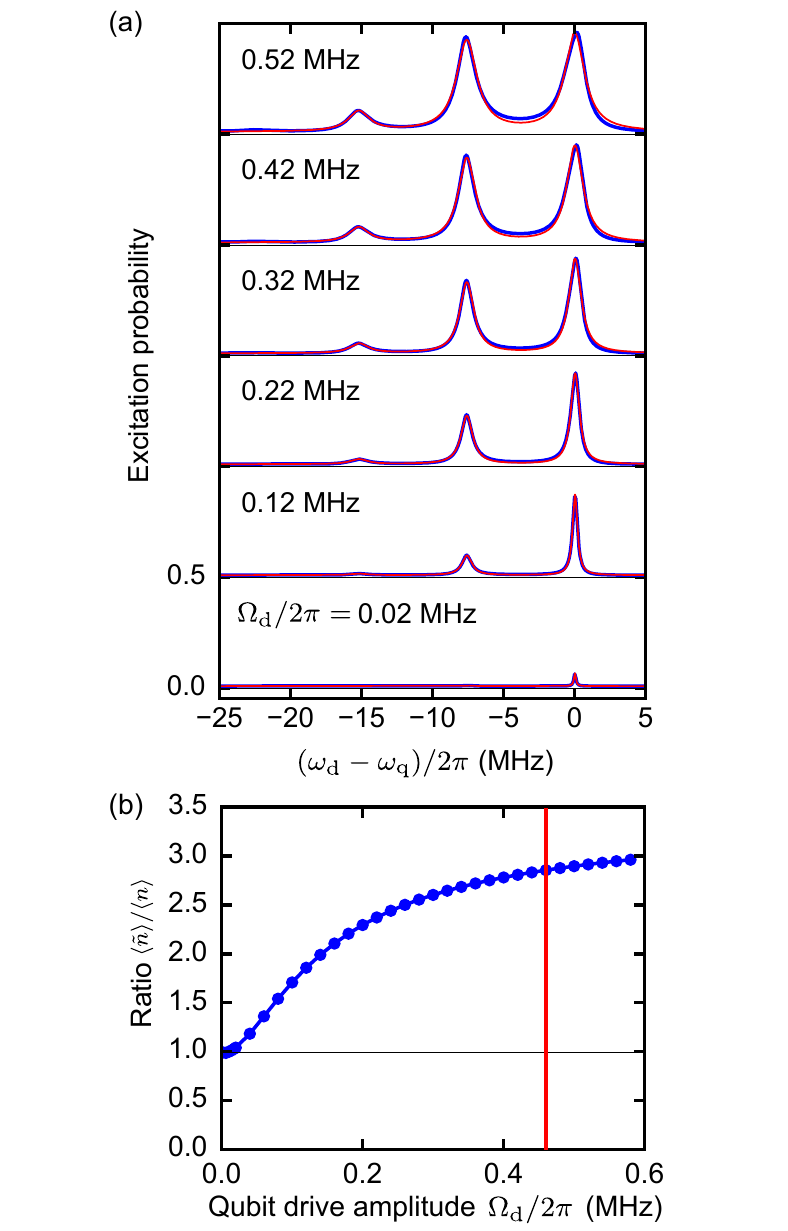} 
\caption{
(a)~Qubit spectrum for each qubit drive amplitude $\Omega_{\rm d}$.
The spectra are offset vertically by 0.5 each. 
Blue solid lines are the numerical results and red solid lines are the multi-Lorentzian fits.
The average photon number in the cavity under a coherent drive is fixed to 0.2.
(b)~Ratio of  the apparent average photon number $\langle \tilde{n}\rangle$ to the actual average photon number $\langle n\rangle$ as a function of the qubit drive amplitude $\Omega_{\rm d}$.
Red solid line indicates the qubit drive amplitude we used in the qubit spectroscopy shown in the main text.
} 
  \label{figs7}
\enc
\end{figure}

In the qubit spectrum for $\omega_{\rm p}=\omega_{\rm c}-\chi$ [Fig.~2(a)], we find a splitting in the peak corresponding to the single photon occupancy. 
The splitting is understood as the Autler-Townes effect involving the three states: $|{\rm e},0\rangle$, $|{\rm e},1\rangle$ and $|{\rm g},1\rangle$ \cite{ate}.
In the cavity probe frequency $\om_{\rm p}$ dependence of the qubit spectra, an anti-crossing like splitting is observed at around $\omega_{\rm p}= \omega_{\rm c}-\chi$ and $\omega_{\rm d} =  \omega_{\rm q} -2\chi$, as shown in the yellow rectangle in Fig.~\ref{figs6}(b) and in Fig.~\ref{figs6}(c).
Due to the thermal photon excitation in the cavity, the population of $|{\rm g},1\rangle$ is finite.
In the steady-state, the qubit drive field at $\omega_{\rm d}=\omega_{\rm q}-2\chi$ transfers the population of $|{\rm g},1\rangle$ to $|{\rm e},0\rangle$,
because the qubit drive amplitude $\Om_{\rm d}$ and the cavity decay rate $\kappa$ are larger than the qubit decay rate $\gamma$.
Therefore, the cavity probe field around $\omega_{\rm p}=\omega_{\rm c}-\chi$ can excite the photons in the cavity from $|{\rm e},0\rangle$.
The qubit drive field couples $|{\rm e},1\rangle$ to $|{\rm g},1\rangle$ and splits the spectrum into the two peaks with the separation of $\Om_{\rm d}$ [see fig.~\ref{figs6}(d)].
The white dashed lines in Fig.~\ref{figs6}(c) depict the transition frequencies from $|{\rm e},0\rangle$ to the hybridized states composed of $|{\rm g},1\rangle$ and $|{\rm e},1\rangle$, which is calculated from the Hamiltonian Eq.~(\ref{hr}) with $\om_{\rm p}=\Om_{\rm p}=0$.

\section*{S5. Effect of qubit drive field}
As shown in Figs.~3 and~4 of the main text, the peak area ratio in the qubit spectrum does not coincide exactly with the actual photon number distribution in the cavity due to the finite qubit drive field.
The discrepancy can be explained by the accumulation of the population in the state $|\rm{e},0\rangle$ in the relation between the excitation and decay rate of the system. 
To be more quantitative, we study the effect of the qubit drive amplitude $\Omega_{\rm d}$ on the qubit spectrum in a numerical simulation setting $S_{\rm A}=(\sigma_z+1)/2$ in Eq.~(\ref{fin}).
As an example, the cavity is driven by a coherent state to have a steady-state with the average photon number 0.2.
Note that the cavity probe power is so weak that the excitation probability of the qubit is proportional to the cavity transmission, which is measured in the experiment.
In Fig.~\ref{figs7}(a), we plot the qubit excitation probability $\langle (\sigma_z + 1)/2\rangle$ as a function of the qubit drive frequency $\omega_{\rm d}$ for each drive amplitude $\Omega_{\rm d}$ (blue solid lines).
As the drive amplitude $\Omega_{\rm d}$ is increased, the peaks in the qubit spectrum are enhanced, which allows us to observe the peaks with a higher signal-to-noise ratio in the experiment.
Especially, this effect makes it easier to characterize a microwave quantum state whose average photon number is small.
The red solid lines are multi-Lorentzian fits to deduce the apparent average photon numbers $\langle \tilde{n}\rangle$ from the peak area ratios.
In Fig.~\ref{figs7}(b), we plot the ratio of $\langle \tilde{n}\rangle$ to the actual average photon number $\langle n\rangle$ as a function of the drive amplitude $\Omega_{\rm d}$.
In the small amplitude limit of the qubit drive, $\langle \tilde{n}\rangle$ is identical to $\langle n\rangle$.
However, as $\Omega_{\rm d}$ increases, the ratio $\langle \tilde{n}\rangle/\langle n\rangle$ increases, meaning that the populations of the larger photon number states are effectively enhanced.

\bibliography{mybib}

\begin{thebibliography}{33}%
\makeatletter
\providecommand \@ifxundefined [1]{%
 \@ifx{#1\undefined}
}%
\providecommand \@ifnum [1]{%
 \ifnum #1\expandafter \@firstoftwo
 \else \expandafter \@secondoftwo
 \fi
}%
\providecommand \@ifx [1]{%
 \ifx #1\expandafter \@firstoftwo
 \else \expandafter \@secondoftwo
 \fi
}%
\providecommand \natexlab [1]{#1}%
\providecommand \enquote  [1]{``#1''}%
\providecommand \bibnamefont  [1]{#1}%
\providecommand \bibfnamefont [1]{#1}%
\providecommand \citenamefont [1]{#1}%
\providecommand \href@noop [0]{\@secondoftwo}%
\providecommand \href [0]{\begingroup \@sanitize@url \@href}%
\providecommand \@href[1]{\@@startlink{#1}\@@href}%
\providecommand \@@href[1]{\endgroup#1\@@endlink}%
\providecommand \@sanitize@url [0]{\catcode `\\12\catcode `\$12\catcode
  `\&12\catcode `\#12\catcode `\^12\catcode `\_12\catcode `\%12\relax}%
\providecommand \@@startlink[1]{}%
\providecommand \@@endlink[0]{}%
\providecommand \url  [0]{\begingroup\@sanitize@url \@url }%
\providecommand \@url [1]{\endgroup\@href {#1}{\urlprefix }}%
\providecommand \urlprefix  [0]{URL }%
\providecommand \Eprint [0]{\href }%
\providecommand \doibase [0]{http://dx.doi.org/}%
\providecommand \selectlanguage [0]{\@gobble}%
\providecommand \bibinfo  [0]{\@secondoftwo}%
\providecommand \bibfield  [0]{\@secondoftwo}%
\providecommand \translation [1]{[#1]}%
\providecommand \BibitemOpen [0]{}%
\providecommand \bibitemStop [0]{}%
\providecommand \bibitemNoStop [0]{.\EOS\space}%
\providecommand \EOS [0]{\spacefactor3000\relax}%
\providecommand \BibitemShut  [1]{\csname bibitem#1\endcsname}%
\let\auto@bib@innerbib\@empty
\bibitem [{\citenamefont {Devoret}\ and\ \citenamefont
  {Schoelkopf}(2013)}]{scq}%
  \BibitemOpen
  \bibfield  {author} {\bibinfo {author} {\bibfnamefont {M.~H.}\ \bibnamefont
  {Devoret}}\ and\ \bibinfo {author} {\bibfnamefont {R.~J.}\ \bibnamefont
  {Schoelkopf}},\ }\href@noop {} {\bibfield  {journal} {\bibinfo  {journal}
  {Science}\ }\textbf {\bibinfo {volume} {339}},\ \bibinfo {pages} {1169}
  (\bibinfo {year} {2013})}\BibitemShut {NoStop}%
\bibitem [{\citenamefont {Blais}\ \emph {et~al.}(2004)\citenamefont {Blais},
  \citenamefont {Huang}, \citenamefont {Wallraff}, \citenamefont {Girvin},\
  and\ \citenamefont {Schoelkopf}}]{cqed}%
  \BibitemOpen
  \bibfield  {author} {\bibinfo {author} {\bibfnamefont {A.}~\bibnamefont
  {Blais}}, \bibinfo {author} {\bibfnamefont {R.-S.}\ \bibnamefont {Huang}},
  \bibinfo {author} {\bibfnamefont {A.}~\bibnamefont {Wallraff}}, \bibinfo
  {author} {\bibfnamefont {S.~M.}\ \bibnamefont {Girvin}}, \ and\ \bibinfo
  {author} {\bibfnamefont {R.~J.}\ \bibnamefont {Schoelkopf}},\ }\href@noop {}
  {\bibfield  {journal} {\bibinfo  {journal} {Phys. Rev. A}\ }\textbf {\bibinfo
  {volume} {69}},\ \bibinfo {pages} {062320} (\bibinfo {year}
  {2004})}\BibitemShut {NoStop}%
\bibitem [{\citenamefont {Drummond}\ and\ \citenamefont {Ficek}(2013)}]{qsq}%
  \BibitemOpen
  \bibfield  {author} {\bibinfo {author} {\bibfnamefont {P.~D.}\ \bibnamefont
  {Drummond}}\ and\ \bibinfo {author} {\bibfnamefont {Z.}~\bibnamefont
  {Ficek}},\ }\href@noop {} {\emph {\bibinfo {title} {Quantum Squeezing}}},\
  Vol.~\bibinfo {volume} {27}\ (\bibinfo  {publisher} {Springer Science \&
  Business Media},\ \bibinfo {year} {2013})\BibitemShut {NoStop}%
\bibitem [{\citenamefont {Movshovich}\ \emph {et~al.}(1990)\citenamefont
  {Movshovich}, \citenamefont {Yurke}, \citenamefont {Kaminsky}, \citenamefont
  {Smith}, \citenamefont {Silver}, \citenamefont {Simon},\ and\ \citenamefont
  {Schneider}}]{ozn}%
  \BibitemOpen
  \bibfield  {author} {\bibinfo {author} {\bibfnamefont {R.}~\bibnamefont
  {Movshovich}}, \bibinfo {author} {\bibfnamefont {B.}~\bibnamefont {Yurke}},
  \bibinfo {author} {\bibfnamefont {P.~G.}\ \bibnamefont {Kaminsky}}, \bibinfo
  {author} {\bibfnamefont {A.~D.}\ \bibnamefont {Smith}}, \bibinfo {author}
  {\bibfnamefont {A.~H.}\ \bibnamefont {Silver}}, \bibinfo {author}
  {\bibfnamefont {R.~W.}\ \bibnamefont {Simon}}, \ and\ \bibinfo {author}
  {\bibfnamefont {M.~V.}\ \bibnamefont {Schneider}},\ }\href@noop {} {\bibfield
   {journal} {\bibinfo  {journal} {Phys. Rev. Lett.}\ }\textbf {\bibinfo
  {volume} {65}},\ \bibinfo {pages} {1419} (\bibinfo {year}
  {1990})}\BibitemShut {NoStop}%
\bibitem [{\citenamefont {Castellanos-Beltran}\ \emph
  {et~al.}(2008)\citenamefont {Castellanos-Beltran}, \citenamefont {Irwin},
  \citenamefont {Hilton}, \citenamefont {Vale},\ and\ \citenamefont
  {Lehnert}}]{asq}%
  \BibitemOpen
  \bibfield  {author} {\bibinfo {author} {\bibfnamefont {M.}~\bibnamefont
  {Castellanos-Beltran}}, \bibinfo {author} {\bibfnamefont {K.}~\bibnamefont
  {Irwin}}, \bibinfo {author} {\bibfnamefont {G.}~\bibnamefont {Hilton}},
  \bibinfo {author} {\bibfnamefont {L.}~\bibnamefont {Vale}}, \ and\ \bibinfo
  {author} {\bibfnamefont {K.}~\bibnamefont {Lehnert}},\ }\href@noop {}
  {\bibfield  {journal} {\bibinfo  {journal} {Nature Phys.}\ }\textbf {\bibinfo
  {volume} {4}},\ \bibinfo {pages} {929} (\bibinfo {year} {2008})}\BibitemShut
  {NoStop}%
\bibitem [{\citenamefont {Mallet}\ \emph {et~al.}(2011)\citenamefont {Mallet},
  \citenamefont {Castellanos-Beltran}, \citenamefont {Ku}, \citenamefont
  {Glancy}, \citenamefont {Knill}, \citenamefont {Irwin}, \citenamefont
  {Hilton}, \citenamefont {Vale},\ and\ \citenamefont {Lehnert}}]{qsm}%
  \BibitemOpen
  \bibfield  {author} {\bibinfo {author} {\bibfnamefont {F.}~\bibnamefont
  {Mallet}}, \bibinfo {author} {\bibfnamefont {M.~A.}\ \bibnamefont
  {Castellanos-Beltran}}, \bibinfo {author} {\bibfnamefont {H.~S.}\
  \bibnamefont {Ku}}, \bibinfo {author} {\bibfnamefont {S.}~\bibnamefont
  {Glancy}}, \bibinfo {author} {\bibfnamefont {E.}~\bibnamefont {Knill}},
  \bibinfo {author} {\bibfnamefont {K.~D.}\ \bibnamefont {Irwin}}, \bibinfo
  {author} {\bibfnamefont {G.~C.}\ \bibnamefont {Hilton}}, \bibinfo {author}
  {\bibfnamefont {L.~R.}\ \bibnamefont {Vale}}, \ and\ \bibinfo {author}
  {\bibfnamefont {K.~W.}\ \bibnamefont {Lehnert}},\ }\href@noop {} {\bibfield
  {journal} {\bibinfo  {journal} {Phys. Rev. Lett.}\ }\textbf {\bibinfo
  {volume} {106}},\ \bibinfo {pages} {220502} (\bibinfo {year}
  {2011})}\BibitemShut {NoStop}%
\bibitem [{\citenamefont {Menzel}\ \emph {et~al.}(2012)\citenamefont {Menzel},
  \citenamefont {Di~Candia}, \citenamefont {Deppe}, \citenamefont {Eder},
  \citenamefont {Zhong}, \citenamefont {Ihmig}, \citenamefont {Haeberlein},
  \citenamefont {Baust}, \citenamefont {Hoffmann}, \citenamefont {Ballester},
  \citenamefont {Inomata}, \citenamefont {Yamamoto}, \citenamefont {Nakamura},
  \citenamefont {Solano}, \citenamefont {Marx},\ and\ \citenamefont
  {Gross}}]{pec}%
  \BibitemOpen
  \bibfield  {author} {\bibinfo {author} {\bibfnamefont {E.~P.}\ \bibnamefont
  {Menzel}}, \bibinfo {author} {\bibfnamefont {R.}~\bibnamefont {Di~Candia}},
  \bibinfo {author} {\bibfnamefont {F.}~\bibnamefont {Deppe}}, \bibinfo
  {author} {\bibfnamefont {P.}~\bibnamefont {Eder}}, \bibinfo {author}
  {\bibfnamefont {L.}~\bibnamefont {Zhong}}, \bibinfo {author} {\bibfnamefont
  {M.}~\bibnamefont {Ihmig}}, \bibinfo {author} {\bibfnamefont
  {M.}~\bibnamefont {Haeberlein}}, \bibinfo {author} {\bibfnamefont
  {A.}~\bibnamefont {Baust}}, \bibinfo {author} {\bibfnamefont
  {E.}~\bibnamefont {Hoffmann}}, \bibinfo {author} {\bibfnamefont
  {D.}~\bibnamefont {Ballester}}, \bibinfo {author} {\bibfnamefont
  {K.}~\bibnamefont {Inomata}}, \bibinfo {author} {\bibfnamefont
  {T.}~\bibnamefont {Yamamoto}}, \bibinfo {author} {\bibfnamefont
  {Y.}~\bibnamefont {Nakamura}}, \bibinfo {author} {\bibfnamefont
  {E.}~\bibnamefont {Solano}}, \bibinfo {author} {\bibfnamefont
  {A.}~\bibnamefont {Marx}}, \ and\ \bibinfo {author} {\bibfnamefont
  {R.}~\bibnamefont {Gross}},\ }\href@noop {} {\bibfield  {journal} {\bibinfo
  {journal} {Phys. Rev. Lett.}\ }\textbf {\bibinfo {volume} {109}},\ \bibinfo
  {pages} {250502} (\bibinfo {year} {2012})}\BibitemShut {NoStop}%
\bibitem [{\citenamefont {Zhong}\ \emph {et~al.}(2013)\citenamefont {Zhong},
  \citenamefont {Menzel}, \citenamefont {Di~Candia}, \citenamefont {Eder},
  \citenamefont {Ihmig}, \citenamefont {Baust}, \citenamefont {Haeberlein},
  \citenamefont {Hoffmann}, \citenamefont {Inomata}, \citenamefont {Yamamoto},
  \citenamefont {Nakamura}, \citenamefont {Solano}, \citenamefont {Deppe},
  \citenamefont {Marx},\ and\ \citenamefont {Gross}}]{sfj}%
  \BibitemOpen
  \bibfield  {author} {\bibinfo {author} {\bibfnamefont {L.}~\bibnamefont
  {Zhong}}, \bibinfo {author} {\bibfnamefont {E.~P.}\ \bibnamefont {Menzel}},
  \bibinfo {author} {\bibfnamefont {R.}~\bibnamefont {Di~Candia}}, \bibinfo
  {author} {\bibfnamefont {P.}~\bibnamefont {Eder}}, \bibinfo {author}
  {\bibfnamefont {M.}~\bibnamefont {Ihmig}}, \bibinfo {author} {\bibfnamefont
  {A.}~\bibnamefont {Baust}}, \bibinfo {author} {\bibfnamefont
  {M.}~\bibnamefont {Haeberlein}}, \bibinfo {author} {\bibfnamefont
  {E.}~\bibnamefont {Hoffmann}}, \bibinfo {author} {\bibfnamefont
  {K.}~\bibnamefont {Inomata}}, \bibinfo {author} {\bibfnamefont
  {T.}~\bibnamefont {Yamamoto}}, \bibinfo {author} {\bibfnamefont
  {Y.}~\bibnamefont {Nakamura}}, \bibinfo {author} {\bibfnamefont
  {E.}~\bibnamefont {Solano}}, \bibinfo {author} {\bibfnamefont
  {F.}~\bibnamefont {Deppe}}, \bibinfo {author} {\bibfnamefont
  {A.}~\bibnamefont {Marx}}, \ and\ \bibinfo {author} {\bibfnamefont
  {R.}~\bibnamefont {Gross}},\ }\href@noop {} {\bibfield  {journal} {\bibinfo
  {journal} {New J. Phys.}\ }\textbf {\bibinfo {volume} {15}},\ \bibinfo
  {pages} {125013} (\bibinfo {year} {2013})}\BibitemShut {NoStop}%
\bibitem [{\citenamefont {Wilson}\ \emph {et~al.}(2011)\citenamefont {Wilson},
  \citenamefont {Johansson}, \citenamefont {Pourkabirian}, \citenamefont
  {Simoen}, \citenamefont {Johansson}, \citenamefont {Duty}, \citenamefont
  {Nori},\ and\ \citenamefont {Delsing}}]{odc}%
  \BibitemOpen
  \bibfield  {author} {\bibinfo {author} {\bibfnamefont {C.~M.}\ \bibnamefont
  {Wilson}}, \bibinfo {author} {\bibfnamefont {G.}~\bibnamefont {Johansson}},
  \bibinfo {author} {\bibfnamefont {A.}~\bibnamefont {Pourkabirian}}, \bibinfo
  {author} {\bibfnamefont {M.}~\bibnamefont {Simoen}}, \bibinfo {author}
  {\bibfnamefont {J.~R.}\ \bibnamefont {Johansson}}, \bibinfo {author}
  {\bibfnamefont {T.}~\bibnamefont {Duty}}, \bibinfo {author} {\bibfnamefont
  {F.}~\bibnamefont {Nori}}, \ and\ \bibinfo {author} {\bibfnamefont
  {P.}~\bibnamefont {Delsing}},\ }\href@noop {} {\bibfield  {journal} {\bibinfo
   {journal} {Nature}\ }\textbf {\bibinfo {volume} {479}},\ \bibinfo {pages}
  {376} (\bibinfo {year} {2011})}\BibitemShut {NoStop}%
\bibitem [{\citenamefont {Eichler}\ \emph {et~al.}(2011)\citenamefont
  {Eichler}, \citenamefont {Bozyigit}, \citenamefont {Lang}, \citenamefont
  {Baur}, \citenamefont {Steffen}, \citenamefont {Fink}, \citenamefont
  {Filipp},\ and\ \citenamefont {Wallraff}}]{ots}%
  \BibitemOpen
  \bibfield  {author} {\bibinfo {author} {\bibfnamefont {C.}~\bibnamefont
  {Eichler}}, \bibinfo {author} {\bibfnamefont {D.}~\bibnamefont {Bozyigit}},
  \bibinfo {author} {\bibfnamefont {C.}~\bibnamefont {Lang}}, \bibinfo {author}
  {\bibfnamefont {M.}~\bibnamefont {Baur}}, \bibinfo {author} {\bibfnamefont
  {L.}~\bibnamefont {Steffen}}, \bibinfo {author} {\bibfnamefont {J.~M.}\
  \bibnamefont {Fink}}, \bibinfo {author} {\bibfnamefont {S.}~\bibnamefont
  {Filipp}}, \ and\ \bibinfo {author} {\bibfnamefont {A.}~\bibnamefont
  {Wallraff}},\ }\href@noop {} {\bibfield  {journal} {\bibinfo  {journal}
  {Phys. Rev. Lett.}\ }\textbf {\bibinfo {volume} {107}},\ \bibinfo {pages}
  {113601} (\bibinfo {year} {2011})}\BibitemShut {NoStop}%
\bibitem [{\citenamefont {Bergeal}\ \emph {et~al.}(2012)\citenamefont
  {Bergeal}, \citenamefont {Schackert}, \citenamefont {Frunzio},\ and\
  \citenamefont {Devoret}}]{tmc}%
  \BibitemOpen
  \bibfield  {author} {\bibinfo {author} {\bibfnamefont {N.}~\bibnamefont
  {Bergeal}}, \bibinfo {author} {\bibfnamefont {F.}~\bibnamefont {Schackert}},
  \bibinfo {author} {\bibfnamefont {L.}~\bibnamefont {Frunzio}}, \ and\
  \bibinfo {author} {\bibfnamefont {M.~H.}\ \bibnamefont {Devoret}},\
  }\href@noop {} {\bibfield  {journal} {\bibinfo  {journal} {Phys. Rev. Lett.}\
  }\textbf {\bibinfo {volume} {108}},\ \bibinfo {pages} {123902} (\bibinfo
  {year} {2012})}\BibitemShut {NoStop}%
\bibitem [{\citenamefont {Flurin}\ \emph {et~al.}(2012)\citenamefont {Flurin},
  \citenamefont {Roch}, \citenamefont {Mallet}, \citenamefont {Devoret},\ and\
  \citenamefont {Huard}}]{gem}%
  \BibitemOpen
  \bibfield  {author} {\bibinfo {author} {\bibfnamefont {E.}~\bibnamefont
  {Flurin}}, \bibinfo {author} {\bibfnamefont {N.}~\bibnamefont {Roch}},
  \bibinfo {author} {\bibfnamefont {F.}~\bibnamefont {Mallet}}, \bibinfo
  {author} {\bibfnamefont {M.~H.}\ \bibnamefont {Devoret}}, \ and\ \bibinfo
  {author} {\bibfnamefont {B.}~\bibnamefont {Huard}},\ }\href@noop {}
  {\bibfield  {journal} {\bibinfo  {journal} {Phys. Rev. Lett.}\ }\textbf
  {\bibinfo {volume} {109}},\ \bibinfo {pages} {183901} (\bibinfo {year}
  {2012})}\BibitemShut {NoStop}%
\bibitem [{\citenamefont {Macklin}\ \emph {et~al.}(2015)\citenamefont
  {Macklin}, \citenamefont {O'Brien}, \citenamefont {Hover}, \citenamefont
  {Schwartz}, \citenamefont {Bolkhovsky}, \citenamefont {Zhang}, \citenamefont
  {Oliver},\ and\ \citenamefont {Siddiqi}}]{jtp}%
  \BibitemOpen
  \bibfield  {author} {\bibinfo {author} {\bibfnamefont {C.}~\bibnamefont
  {Macklin}}, \bibinfo {author} {\bibfnamefont {K.}~\bibnamefont {O'Brien}},
  \bibinfo {author} {\bibfnamefont {D.}~\bibnamefont {Hover}}, \bibinfo
  {author} {\bibfnamefont {M.~E.}\ \bibnamefont {Schwartz}}, \bibinfo {author}
  {\bibfnamefont {V.}~\bibnamefont {Bolkhovsky}}, \bibinfo {author}
  {\bibfnamefont {X.}~\bibnamefont {Zhang}}, \bibinfo {author} {\bibfnamefont
  {W.~D.}\ \bibnamefont {Oliver}}, \ and\ \bibinfo {author} {\bibfnamefont
  {I.}~\bibnamefont {Siddiqi}},\ }\href@noop {} {\bibfield  {journal} {\bibinfo
   {journal} {Science}\ }\textbf {\bibinfo {volume} {350}},\ \bibinfo {pages}
  {307} (\bibinfo {year} {2015})}\BibitemShut {NoStop}%
\bibitem [{\citenamefont {Murch}\ \emph {et~al.}(2013)\citenamefont {Murch},
  \citenamefont {Weber}, \citenamefont {Beck}, \citenamefont {Ginossar},\ and\
  \citenamefont {Siddiqi}}]{rrd}%
  \BibitemOpen
  \bibfield  {author} {\bibinfo {author} {\bibfnamefont {K.~W.}\ \bibnamefont
  {Murch}}, \bibinfo {author} {\bibfnamefont {S.~J.}\ \bibnamefont {Weber}},
  \bibinfo {author} {\bibfnamefont {K.~M.}\ \bibnamefont {Beck}}, \bibinfo
  {author} {\bibfnamefont {E.}~\bibnamefont {Ginossar}}, \ and\ \bibinfo
  {author} {\bibfnamefont {I.}~\bibnamefont {Siddiqi}},\ }\href@noop {}
  {\bibfield  {journal} {\bibinfo  {journal} {Nature}\ }\textbf {\bibinfo
  {volume} {499}},\ \bibinfo {pages} {62} (\bibinfo {year} {2013})}\BibitemShut
  {NoStop}%
\bibitem [{\citenamefont {Toyli}\ \emph {et~al.}(2016)\citenamefont {Toyli},
  \citenamefont {Eddins}, \citenamefont {Boutin}, \citenamefont {Puri},
  \citenamefont {Hover}, \citenamefont {Bolkhovsky}, \citenamefont {Oliver},
  \citenamefont {Blais},\ and\ \citenamefont {Siddiqi}}]{rfa}%
  \BibitemOpen
  \bibfield  {author} {\bibinfo {author} {\bibfnamefont {D.~M.}\ \bibnamefont
  {Toyli}}, \bibinfo {author} {\bibfnamefont {A.~W.}\ \bibnamefont {Eddins}},
  \bibinfo {author} {\bibfnamefont {S.}~\bibnamefont {Boutin}}, \bibinfo
  {author} {\bibfnamefont {S.}~\bibnamefont {Puri}}, \bibinfo {author}
  {\bibfnamefont {D.}~\bibnamefont {Hover}}, \bibinfo {author} {\bibfnamefont
  {V.}~\bibnamefont {Bolkhovsky}}, \bibinfo {author} {\bibfnamefont {W.~D.}\
  \bibnamefont {Oliver}}, \bibinfo {author} {\bibfnamefont {A.}~\bibnamefont
  {Blais}}, \ and\ \bibinfo {author} {\bibfnamefont {I.}~\bibnamefont
  {Siddiqi}},\ }\href@noop {} {\bibfield  {journal} {\bibinfo  {journal} {Phys.
  Rev. X}\ }\textbf {\bibinfo {volume} {6}},\ \bibinfo {pages} {031004}
  (\bibinfo {year} {2016})}\BibitemShut {NoStop}%
\bibitem [{\citenamefont {Bienfait}\ \emph {et~al.}(2016)\citenamefont
  {Bienfait}, \citenamefont {Campagne-Ibarcq}, \citenamefont {Holm-Kiilerich},
  \citenamefont {Zhou}, \citenamefont {Probst}, \citenamefont {Pla},
  \citenamefont {Schenkel}, \citenamefont {Vion}, \citenamefont {Esteve},
  \citenamefont {Morton}, \citenamefont {Moelmer},\ and\ \citenamefont
  {Bertet}}]{mrs}%
  \BibitemOpen
  \bibfield  {author} {\bibinfo {author} {\bibfnamefont {A.}~\bibnamefont
  {Bienfait}}, \bibinfo {author} {\bibfnamefont {P.}~\bibnamefont
  {Campagne-Ibarcq}}, \bibinfo {author} {\bibfnamefont {A.}~\bibnamefont
  {Holm-Kiilerich}}, \bibinfo {author} {\bibfnamefont {X.}~\bibnamefont
  {Zhou}}, \bibinfo {author} {\bibfnamefont {S.}~\bibnamefont {Probst}},
  \bibinfo {author} {\bibfnamefont {J.~J.}\ \bibnamefont {Pla}}, \bibinfo
  {author} {\bibfnamefont {T.}~\bibnamefont {Schenkel}}, \bibinfo {author}
  {\bibfnamefont {D.}~\bibnamefont {Vion}}, \bibinfo {author} {\bibfnamefont
  {D.}~\bibnamefont {Esteve}}, \bibinfo {author} {\bibfnamefont {J.~J.~L.}\
  \bibnamefont {Morton}}, \bibinfo {author} {\bibfnamefont {K.}~\bibnamefont
  {Moelmer}}, \ and\ \bibinfo {author} {\bibfnamefont {P.}~\bibnamefont
  {Bertet}},\ }\href@noop {} {\bibfield  {journal} {\bibinfo  {journal}
  {arXiv:1610.03329}\ } (\bibinfo {year} {2016})}\BibitemShut {NoStop}%
\bibitem [{\citenamefont {Schleich}\ and\ \citenamefont {Wheeler}(1987)}]{eoo}%
  \BibitemOpen
  \bibfield  {author} {\bibinfo {author} {\bibfnamefont {W.}~\bibnamefont
  {Schleich}}\ and\ \bibinfo {author} {\bibfnamefont {J.}~\bibnamefont
  {Wheeler}},\ }\href@noop {} {\bibfield  {journal} {\bibinfo  {journal}
  {Nature}\ }\textbf {\bibinfo {volume} {326}},\ \bibinfo {pages} {574}
  (\bibinfo {year} {1987})}\BibitemShut {NoStop}%
\bibitem [{\citenamefont {Waks}\ \emph {et~al.}(2006)\citenamefont {Waks},
  \citenamefont {Sanders}, \citenamefont {Diamanti},\ and\ \citenamefont
  {Yamamoto}}]{hnp}%
  \BibitemOpen
  \bibfield  {author} {\bibinfo {author} {\bibfnamefont {E.}~\bibnamefont
  {Waks}}, \bibinfo {author} {\bibfnamefont {B.~C.}\ \bibnamefont {Sanders}},
  \bibinfo {author} {\bibfnamefont {E.}~\bibnamefont {Diamanti}}, \ and\
  \bibinfo {author} {\bibfnamefont {Y.}~\bibnamefont {Yamamoto}},\ }\href@noop
  {} {\bibfield  {journal} {\bibinfo  {journal} {Phys. Rev. A}\ }\textbf
  {\bibinfo {volume} {73}},\ \bibinfo {pages} {033814} (\bibinfo {year}
  {2006})}\BibitemShut {NoStop}%
\bibitem [{\citenamefont {Wakui}\ \emph {et~al.}(2014)\citenamefont {Wakui},
  \citenamefont {Eto}, \citenamefont {Benichi}, \citenamefont {Izumi},
  \citenamefont {Yanagida}, \citenamefont {Ema}, \citenamefont {Numata},
  \citenamefont {Fukuda}, \citenamefont {Takeoka},\ and\ \citenamefont
  {Sasaki}}]{udd}%
  \BibitemOpen
  \bibfield  {author} {\bibinfo {author} {\bibfnamefont {K.}~\bibnamefont
  {Wakui}}, \bibinfo {author} {\bibfnamefont {Y.}~\bibnamefont {Eto}}, \bibinfo
  {author} {\bibfnamefont {H.}~\bibnamefont {Benichi}}, \bibinfo {author}
  {\bibfnamefont {S.}~\bibnamefont {Izumi}}, \bibinfo {author} {\bibfnamefont
  {T.}~\bibnamefont {Yanagida}}, \bibinfo {author} {\bibfnamefont
  {K.}~\bibnamefont {Ema}}, \bibinfo {author} {\bibfnamefont {T.}~\bibnamefont
  {Numata}}, \bibinfo {author} {\bibfnamefont {D.}~\bibnamefont {Fukuda}},
  \bibinfo {author} {\bibfnamefont {M.}~\bibnamefont {Takeoka}}, \ and\
  \bibinfo {author} {\bibfnamefont {M.}~\bibnamefont {Sasaki}},\ }\href@noop {}
  {\bibfield  {journal} {\bibinfo  {journal} {Sci. Rep.}\ }\textbf {\bibinfo
  {volume} {4}},\ \bibinfo {pages} {4535} (\bibinfo {year} {2014})}\BibitemShut
  {NoStop}%
\bibitem [{\citenamefont {Chen}\ \emph {et~al.}(2011)\citenamefont {Chen},
  \citenamefont {Hover}, \citenamefont {Sendelbach}, \citenamefont {Maurer},
  \citenamefont {Merkel}, \citenamefont {Pritchett}, \citenamefont {Wilhelm},\
  and\ \citenamefont {McDermott}}]{mpc}%
  \BibitemOpen
  \bibfield  {author} {\bibinfo {author} {\bibfnamefont {Y.-F.}\ \bibnamefont
  {Chen}}, \bibinfo {author} {\bibfnamefont {D.}~\bibnamefont {Hover}},
  \bibinfo {author} {\bibfnamefont {S.}~\bibnamefont {Sendelbach}}, \bibinfo
  {author} {\bibfnamefont {L.}~\bibnamefont {Maurer}}, \bibinfo {author}
  {\bibfnamefont {S.~T.}\ \bibnamefont {Merkel}}, \bibinfo {author}
  {\bibfnamefont {E.~J.}\ \bibnamefont {Pritchett}}, \bibinfo {author}
  {\bibfnamefont {F.~K.}\ \bibnamefont {Wilhelm}}, \ and\ \bibinfo {author}
  {\bibfnamefont {R.}~\bibnamefont {McDermott}},\ }\href@noop {} {\bibfield
  {journal} {\bibinfo  {journal} {Phys. Rev. Lett}\ }\textbf {\bibinfo {volume}
  {107}},\ \bibinfo {pages} {217401} (\bibinfo {year} {2011})}\BibitemShut
  {NoStop}%
\bibitem [{\citenamefont {Inomata}\ \emph {et~al.}(2016)\citenamefont
  {Inomata}, \citenamefont {Lin}, \citenamefont {Koshino}, \citenamefont
  {Oliver}, \citenamefont {Tsai}, \citenamefont {Yamamoto},\ and\ \citenamefont
  {Nakamura}}]{spd}%
  \BibitemOpen
  \bibfield  {author} {\bibinfo {author} {\bibfnamefont {K.}~\bibnamefont
  {Inomata}}, \bibinfo {author} {\bibfnamefont {Z.}~\bibnamefont {Lin}},
  \bibinfo {author} {\bibfnamefont {K.}~\bibnamefont {Koshino}}, \bibinfo
  {author} {\bibfnamefont {W.~D.}\ \bibnamefont {Oliver}}, \bibinfo {author}
  {\bibfnamefont {J.~S.}\ \bibnamefont {Tsai}}, \bibinfo {author}
  {\bibfnamefont {T.}~\bibnamefont {Yamamoto}}, \ and\ \bibinfo {author}
  {\bibfnamefont {Y.}~\bibnamefont {Nakamura}},\ }\href@noop {} {\bibfield
  {journal} {\bibinfo  {journal} {Nature Commun.}\ }\textbf {\bibinfo {volume}
  {7}},\ \bibinfo {pages} {12303} (\bibinfo {year} {2016})}\BibitemShut
  {NoStop}%
\bibitem [{\citenamefont {Narla}\ \emph {et~al.}(2016)\citenamefont {Narla},
  \citenamefont {Shankar}, \citenamefont {Hatridge}, \citenamefont {Leghtas},
  \citenamefont {Sliwa}, \citenamefont {Zalys-Geller}, \citenamefont
  {Mundhada}, \citenamefont {Pfaff}, \citenamefont {Frunzio}, \citenamefont
  {Schoelkopf},\ and\ \citenamefont {Devoret}}]{rcr}%
  \BibitemOpen
  \bibfield  {author} {\bibinfo {author} {\bibfnamefont {A.}~\bibnamefont
  {Narla}}, \bibinfo {author} {\bibfnamefont {S.}~\bibnamefont {Shankar}},
  \bibinfo {author} {\bibfnamefont {M.}~\bibnamefont {Hatridge}}, \bibinfo
  {author} {\bibfnamefont {Z.}~\bibnamefont {Leghtas}}, \bibinfo {author}
  {\bibfnamefont {K.~M.}\ \bibnamefont {Sliwa}}, \bibinfo {author}
  {\bibfnamefont {E.}~\bibnamefont {Zalys-Geller}}, \bibinfo {author}
  {\bibfnamefont {S.~O.}\ \bibnamefont {Mundhada}}, \bibinfo {author}
  {\bibfnamefont {W.}~\bibnamefont {Pfaff}}, \bibinfo {author} {\bibfnamefont
  {L.}~\bibnamefont {Frunzio}}, \bibinfo {author} {\bibfnamefont {R.~J.}\
  \bibnamefont {Schoelkopf}}, \ and\ \bibinfo {author} {\bibfnamefont {M.~H.}\
  \bibnamefont {Devoret}},\ }\href@noop {} {\bibfield  {journal} {\bibinfo
  {journal} {Phys. Rev. X}\ }\textbf {\bibinfo {volume} {6}},\ \bibinfo {pages}
  {031036} (\bibinfo {year} {2016})}\BibitemShut {NoStop}%
\bibitem [{\citenamefont {Schuster}\ \emph {et~al.}(2007)\citenamefont
  {Schuster}, \citenamefont {Houck}, \citenamefont {Schreier}, \citenamefont
  {Wallraff}, \citenamefont {Gambetta}, \citenamefont {Blais}, \citenamefont
  {Frunzio}, \citenamefont {Majer}, \citenamefont {Johnson}, \citenamefont
  {Devoret}, \citenamefont {Girvin},\ and\ \citenamefont {Schoelkopf}}]{rpn}%
  \BibitemOpen
  \bibfield  {author} {\bibinfo {author} {\bibfnamefont {D.~I.}\ \bibnamefont
  {Schuster}}, \bibinfo {author} {\bibfnamefont {A.~A.}\ \bibnamefont {Houck}},
  \bibinfo {author} {\bibfnamefont {J.~A.}\ \bibnamefont {Schreier}}, \bibinfo
  {author} {\bibfnamefont {A.}~\bibnamefont {Wallraff}}, \bibinfo {author}
  {\bibfnamefont {J.~M.}\ \bibnamefont {Gambetta}}, \bibinfo {author}
  {\bibfnamefont {A.}~\bibnamefont {Blais}}, \bibinfo {author} {\bibfnamefont
  {L.}~\bibnamefont {Frunzio}}, \bibinfo {author} {\bibfnamefont
  {J.}~\bibnamefont {Majer}}, \bibinfo {author} {\bibfnamefont
  {B.}~\bibnamefont {Johnson}}, \bibinfo {author} {\bibfnamefont {M.~H.}\
  \bibnamefont {Devoret}}, \bibinfo {author} {\bibfnamefont {S.~H.}\
  \bibnamefont {Girvin}}, \ and\ \bibinfo {author} {\bibfnamefont {R.~J.}\
  \bibnamefont {Schoelkopf}},\ }\href@noop {} {\bibfield  {journal} {\bibinfo
  {journal} {Nature}\ }\textbf {\bibinfo {volume} {445}},\ \bibinfo {pages}
  {515} (\bibinfo {year} {2007})}\BibitemShut {NoStop}%
\bibitem [{\citenamefont {Suri}\ \emph {et~al.}(2015)\citenamefont {Suri},
  \citenamefont {Keane}, \citenamefont {Bishop}, \citenamefont {Novikov},
  \citenamefont {Wellstood},\ and\ \citenamefont {Palmer}}]{nmp}%
  \BibitemOpen
  \bibfield  {author} {\bibinfo {author} {\bibfnamefont {B.}~\bibnamefont
  {Suri}}, \bibinfo {author} {\bibfnamefont {Z.~K.}\ \bibnamefont {Keane}},
  \bibinfo {author} {\bibfnamefont {L.~S.}\ \bibnamefont {Bishop}}, \bibinfo
  {author} {\bibfnamefont {S.}~\bibnamefont {Novikov}}, \bibinfo {author}
  {\bibfnamefont {F.~C.}\ \bibnamefont {Wellstood}}, \ and\ \bibinfo {author}
  {\bibfnamefont {B.~S.}\ \bibnamefont {Palmer}},\ }\href@noop {} {\bibfield
  {journal} {\bibinfo  {journal} {Phys. Rev. A}\ }\textbf {\bibinfo {volume}
  {92}},\ \bibinfo {pages} {063801} (\bibinfo {year} {2015})}\BibitemShut
  {NoStop}%
\bibitem [{\citenamefont {Gambetta}\ \emph {et~al.}(2006)\citenamefont
  {Gambetta}, \citenamefont {Blais}, \citenamefont {Schuster}, \citenamefont
  {Wallraff}, \citenamefont {Frunzio}, \citenamefont {Majer}, \citenamefont
  {Devoret}, \citenamefont {Girvin},\ and\ \citenamefont {Schoelkopf}}]{qip}%
  \BibitemOpen
  \bibfield  {author} {\bibinfo {author} {\bibfnamefont {J.}~\bibnamefont
  {Gambetta}}, \bibinfo {author} {\bibfnamefont {A.}~\bibnamefont {Blais}},
  \bibinfo {author} {\bibfnamefont {D.~I.}\ \bibnamefont {Schuster}}, \bibinfo
  {author} {\bibfnamefont {A.}~\bibnamefont {Wallraff}}, \bibinfo {author}
  {\bibfnamefont {L.}~\bibnamefont {Frunzio}}, \bibinfo {author} {\bibfnamefont
  {J.}~\bibnamefont {Majer}}, \bibinfo {author} {\bibfnamefont {M.~H.}\
  \bibnamefont {Devoret}}, \bibinfo {author} {\bibfnamefont {S.~M.}\
  \bibnamefont {Girvin}}, \ and\ \bibinfo {author} {\bibfnamefont {R.~J.}\
  \bibnamefont {Schoelkopf}},\ }\href@noop {} {\bibfield  {journal} {\bibinfo
  {journal} {Phys. Rev. A}\ }\textbf {\bibinfo {volume} {74}},\ \bibinfo
  {pages} {042318} (\bibinfo {year} {2006})}\BibitemShut {NoStop}%
\bibitem [{\citenamefont {Klyshko}(1996)}]{kly}%
  \BibitemOpen
  \bibfield  {author} {\bibinfo {author} {\bibfnamefont {D.~N.}\ \bibnamefont
  {Klyshko}},\ }\href@noop {} {\bibfield  {journal} {\bibinfo  {journal} {Phys.
  Lett. A}\ }\textbf {\bibinfo {volume} {213}},\ \bibinfo {pages} {7} (\bibinfo
  {year} {1996})}\BibitemShut {NoStop}%
\bibitem [{\citenamefont {Clerk}\ \emph {et~al.}(2010)\citenamefont {Clerk},
  \citenamefont {Devoret}, \citenamefont {Girvin}, \citenamefont {Marquardt},\
  and\ \citenamefont {Schoelkopf}}]{iqn}%
  \BibitemOpen
  \bibfield  {author} {\bibinfo {author} {\bibfnamefont {A.~A.}\ \bibnamefont
  {Clerk}}, \bibinfo {author} {\bibfnamefont {M.~H.}\ \bibnamefont {Devoret}},
  \bibinfo {author} {\bibfnamefont {S.~M.}\ \bibnamefont {Girvin}}, \bibinfo
  {author} {\bibfnamefont {F.}~\bibnamefont {Marquardt}}, \ and\ \bibinfo
  {author} {\bibfnamefont {R.~J.}\ \bibnamefont {Schoelkopf}},\ }\href@noop {}
  {\bibfield  {journal} {\bibinfo  {journal} {Rev. Mod. Phys.}\ }\textbf
  {\bibinfo {volume} {82}},\ \bibinfo {pages} {1155} (\bibinfo {year}
  {2010})}\BibitemShut {NoStop}%
\bibitem [{\citenamefont {Hofheinz}\ \emph {et~al.}(2009)\citenamefont
  {Hofheinz}, \citenamefont {Wang}, \citenamefont {Ansmann}, \citenamefont
  {Bialczak}, \citenamefont {Lucero}, \citenamefont {Neeley}, \citenamefont
  {O'Connell}, \citenamefont {Sank}, \citenamefont {Wenner}, \citenamefont
  {Martinis},\ and\ \citenamefont {Cleland}}]{saq}%
  \BibitemOpen
  \bibfield  {author} {\bibinfo {author} {\bibfnamefont {M.}~\bibnamefont
  {Hofheinz}}, \bibinfo {author} {\bibfnamefont {H.}~\bibnamefont {Wang}},
  \bibinfo {author} {\bibfnamefont {M.}~\bibnamefont {Ansmann}}, \bibinfo
  {author} {\bibfnamefont {R.~C.}\ \bibnamefont {Bialczak}}, \bibinfo {author}
  {\bibfnamefont {E.}~\bibnamefont {Lucero}}, \bibinfo {author} {\bibfnamefont
  {M.}~\bibnamefont {Neeley}}, \bibinfo {author} {\bibfnamefont {A.~D.}\
  \bibnamefont {O'Connell}}, \bibinfo {author} {\bibfnamefont {D.}~\bibnamefont
  {Sank}}, \bibinfo {author} {\bibfnamefont {J.}~\bibnamefont {Wenner}},
  \bibinfo {author} {\bibfnamefont {J.~M.}\ \bibnamefont {Martinis}}, \ and\
  \bibinfo {author} {\bibfnamefont {A.~N.}\ \bibnamefont {Cleland}},\
  }\href@noop {} {\bibfield  {journal} {\bibinfo  {journal} {Nature}\ }\textbf
  {\bibinfo {volume} {459}},\ \bibinfo {pages} {546} (\bibinfo {year}
  {2009})}\BibitemShut {NoStop}%
\bibitem [{\citenamefont {Vlastakis}\ \emph {et~al.}(2013)\citenamefont
  {Vlastakis}, \citenamefont {Kirchmair}, \citenamefont {Leghtas},
  \citenamefont {Nigg}, \citenamefont {Frunzio}, \citenamefont {Girvin},
  \citenamefont {Mirrahimi}, \citenamefont {Devoret},\ and\ \citenamefont
  {Schoelkopf}}]{deq}%
  \BibitemOpen
  \bibfield  {author} {\bibinfo {author} {\bibfnamefont {B.}~\bibnamefont
  {Vlastakis}}, \bibinfo {author} {\bibfnamefont {G.}~\bibnamefont
  {Kirchmair}}, \bibinfo {author} {\bibfnamefont {Z.}~\bibnamefont {Leghtas}},
  \bibinfo {author} {\bibfnamefont {S.~E.}\ \bibnamefont {Nigg}}, \bibinfo
  {author} {\bibfnamefont {L.}~\bibnamefont {Frunzio}}, \bibinfo {author}
  {\bibfnamefont {S.~M.}\ \bibnamefont {Girvin}}, \bibinfo {author}
  {\bibfnamefont {M.}~\bibnamefont {Mirrahimi}}, \bibinfo {author}
  {\bibfnamefont {M.~H.}\ \bibnamefont {Devoret}}, \ and\ \bibinfo {author}
  {\bibfnamefont {R.~J.}\ \bibnamefont {Schoelkopf}},\ }\href@noop {}
  {\bibfield  {journal} {\bibinfo  {journal} {Science}\ }\textbf {\bibinfo
  {volume} {342}},\ \bibinfo {pages} {607} (\bibinfo {year}
  {2013})}\BibitemShut {NoStop}%
\bibitem [{Sup()}]{Supple}%
  \BibitemOpen
  \href@noop {} {\bibinfo  {journal} {See accompanying supplementary materials
  for details}\ }\BibitemShut {NoStop}%
\bibitem [{\citenamefont {Yamamoto}\ \emph {et~al.}(2008)\citenamefont
  {Yamamoto}, \citenamefont {Inomata}, \citenamefont {Watanabe}, \citenamefont
  {Matsuba}, \citenamefont {Miyazaki}, \citenamefont {Oliver}, \citenamefont
  {Nakamura},\ and\ \citenamefont {Tsai}}]{fdj}%
  \BibitemOpen
\bibfield  {journal} {  }\bibfield  {author} {\bibinfo {author} {\bibfnamefont
  {T.}~\bibnamefont {Yamamoto}}, \bibinfo {author} {\bibfnamefont
  {K.}~\bibnamefont {Inomata}}, \bibinfo {author} {\bibfnamefont
  {M.}~\bibnamefont {Watanabe}}, \bibinfo {author} {\bibfnamefont
  {K.}~\bibnamefont {Matsuba}}, \bibinfo {author} {\bibfnamefont
  {T.}~\bibnamefont {Miyazaki}}, \bibinfo {author} {\bibfnamefont
  {W.}~\bibnamefont {Oliver}}, \bibinfo {author} {\bibfnamefont
  {Y.}~\bibnamefont {Nakamura}}, \ and\ \bibinfo {author} {\bibfnamefont
  {J.}~\bibnamefont {Tsai}},\ }\href@noop {} {\bibfield  {journal} {\bibinfo
  {journal} {Appl. Phys. Lett.}\ }\textbf {\bibinfo {volume} {93}},\ \bibinfo
  {pages} {042510} (\bibinfo {year} {2008})}\BibitemShut {NoStop}%
\bibitem [{\citenamefont {Clerk}\ and\ \citenamefont {Utami}(2007)}]{qth}%
  \BibitemOpen
  \bibfield  {author} {\bibinfo {author} {\bibfnamefont {A.~A.}\ \bibnamefont
  {Clerk}}\ and\ \bibinfo {author} {\bibfnamefont {D.~W.}\ \bibnamefont
  {Utami}},\ }\href@noop {} {\bibfield  {journal} {\bibinfo  {journal} {Phys.
  Rev. A}\ }\textbf {\bibinfo {volume} {75}},\ \bibinfo {pages} {042302}
  (\bibinfo {year} {2007})}\BibitemShut {NoStop}%
\bibitem [{\citenamefont {Suri}\ \emph {et~al.}(2013)\citenamefont {Suri},
  \citenamefont {Keane}, \citenamefont {Ruskov}, \citenamefont {Bishop},
  \citenamefont {Tahan}, \citenamefont {Novikov}, \citenamefont {Robinson},
  \citenamefont {Wellstood},\ and\ \citenamefont {Palmer}}]{ate}%
  \BibitemOpen
  \bibfield  {author} {\bibinfo {author} {\bibfnamefont {B.}~\bibnamefont
  {Suri}}, \bibinfo {author} {\bibfnamefont {Z.~K.}\ \bibnamefont {Keane}},
  \bibinfo {author} {\bibfnamefont {R.}~\bibnamefont {Ruskov}}, \bibinfo
  {author} {\bibfnamefont {L.~S.}\ \bibnamefont {Bishop}}, \bibinfo {author}
  {\bibfnamefont {C.}~\bibnamefont {Tahan}}, \bibinfo {author} {\bibfnamefont
  {S.}~\bibnamefont {Novikov}}, \bibinfo {author} {\bibfnamefont {J.~E.}\
  \bibnamefont {Robinson}}, \bibinfo {author} {\bibfnamefont {F.~C.}\
  \bibnamefont {Wellstood}}, \ and\ \bibinfo {author} {\bibfnamefont {B.~S.}\
  \bibnamefont {Palmer}},\ }\href@noop {} {\bibfield  {journal} {\bibinfo
  {journal} {New J. of Phys.}\ }\textbf {\bibinfo {volume} {15}},\ \bibinfo
  {pages} {125007} (\bibinfo {year} {2013})}\BibitemShut {NoStop}%
\end{thebibliography}%

\end{document}